\documentclass[10pt]{article}
\usepackage{pdfpages}
\usepackage{float}
\usepackage{amsmath}
\usepackage{mathtools}
\usepackage{amsfonts}
\usepackage{jheppub}
\usepackage{color}
\relax
\def\Higgs{h}
\def\qb{\bar{q}}
\def\x{\times}
\def\detS{\left|S_{1\x2\x3\x4}\right|}
\def\mC{\mathcal{C}}
\def\tr{{\rm tr}}
\def\DeltaFour{G}
\def\spa#1.#2{\left\langle#1\,#2\right\rangle}
\def\spb#1.#2{\left[#1\,#2\right]}
\def\spab#1.#2.#3{\left\langle#1|#2|#3\right]}
\def\spba#1.#2.#3{\left[#1|#2|#3\right\rangle}
\def\cH{{\cal H}}
\def\cA{{\cal A}}
\def\slsh{\rlap{$\;\!\!\not$}}     
\def\TrfourgL#1.#2.#3.#4{{\rm tr}_{-}\{{#1}\,{#2}\,{#3}\,{#4}\}}
\def\TrfourgR#1.#2.#3.#4{{\rm tr}_{+}\{{#1}\,{#2}\,{#3}\,{#4}\}}
\def\Trfour#1.#2.#3.#4{{\rm tr}\{{#1}\,{#2}\,{#3}\,{#4}\}}
\def\Trtxudxqxud{\TrfourgL{\slsh{p}_3}.{\slsh{p}_{12}}.{\slsh{p}_4}.{\slsh{p}_{12}}}
\def\sp{\slsh{p}}
\def\ehat{\hat{e}}
\def\Trfourgfive#1.#2.#3.#4{{\rm tr}_5\{{#1}\,{#2}\,{#3}\,{#4}\}}
\def\symbrack{}
\def\DeltaThree{\Delta_3}
\def\tree{{\rm tree}}
\def\ellb{\bar{\ell}}
\def\jb{\bar{j}}
\def\spaa#1.#2.#3.#4{\left\langle#1|#2|#3|#4\right\rangle}
\def\spbb#1.#2.#3.#4{\left[#1|#2|#3|#4\right]}
\def\beq{\begin{equation}}
\def\eeq{\end{equation}}
\def\beqn{\begin{eqnarray}}
\def\eeqn{\end{eqnarray}}

\title{The one-loop amplitudes for Higgs + 4 partons with full mass effects}

\author[a]{Lucy Budge,}
\emailAdd{lucy.budge@durham.ac.uk}
  
\author[b]{John M. Campbell,}
\emailAdd{johnmc@fnal.gov}

\author[a]{Giuseppe De Laurentis,}
\emailAdd{giuseppe.de-laurentis@durham.ac.uk}

\author[a]{R. Keith Ellis,}
\emailAdd{keith.ellis@durham.ac.uk}

\author[c]{Satyajit Seth}
\emailAdd{seth@prl.res.in}

\affiliation[a]{Institute for Particle Physics Phenomenology, Durham University, Durham, DH1 3LE, UK}
\affiliation[b]{Fermilab, PO Box 500, Batavia IL 60510-5011, USA}
\affiliation[c]{Physical Research Laboratory, Navrangpura, Ahmedabad - 380009, India}

\date{\today}

\preprint{FERMILAB-PUB-20-062-T,\, IPPP/20/3} 

\abstract{We present compact analytic formulae for the one-loop
amplitudes for Higgs + 4 parton scattering, $0 \to g g g g \Higgs$,
$0 \to \qb q gg \Higgs$ and $0\to \qb q \qb^\prime q^\prime \Higgs$,
mediated by a loop of massive coloured quarks. We exploit the
correspondence with a theory in which a massive coloured scalar
circulates in the loop to avoid a proliferation in the number of
terms in the result.  In addition, we use momentum twistors and high
precision numerical evaluations to simplify the expressions.  The
analytic results in this paper, in terms of spinor products, allow
construction of an efficient numerical program to calculate the amplitude. }


\begin{document}

\maketitle

\section{Introduction}
At the Large Hadron Collider (LHC) the primary mechanism for producing and detecting Higgs bosons is
the process $gg \to \Higgs$.  This process is mediated, in the Standard Model, by a loop of massive
coloured fermions.  Since the Yukawa coupling is proportional to the fermion mass the predominant contribution
is the result of the coupling of the top quark to the Higgs boson.  In the limit in which only a very heavy top quark contributes,
the corresponding amplitude is independent of the top quark mass;  this gives rise to an effective field
theory (EFT) in which the loop of heavy top quarks is replaced by an effective Lagrangian,
\begin{equation}
\label{EFT}
{\mathcal L}_{{\rm eff}} = \frac{g_s^2}{48\pi^2 v} \, \Higgs \, G_{\mu\nu}^A G^{A,\mu\nu}
\end{equation}
where $g_s$ is the strong coupling constant, $v$ is the vacuum expectation value of the Higgs field,
$G_{\mu\nu}$ is QCD field strength, and $\Higgs$ is the Higgs boson field.
The EFT in Eq.~(\ref{EFT}) has
been used to compute higher-order corrections to the inclusive cross-section -- most recently up
to next-to-next-to-next-to-leading order~\cite{Anastasiou:2015ema,Mistlberger:2018etf}
 -- as well as rates for the production of Higgs bosons in
association with up to three additional jets up to next-to-leading
order~\cite{Campbell:2006xx,Cullen:2013saa}.  The effective field theory description is expected 
to break down when, for example, the transverse momentum of produced gluons is of order of the 
top quark mass. This breakdown has most recently been investigated at NLO in ref.~\cite{Jones:2018hbb}.  
This kinematic regime is beginning to be explored at the LHC~\cite{Sirunyan:2017dgc} and can give important
information about the mediators in the loop that couple to the Higgs.  For such configurations it
is therefore important to make use of a superior calculation in which the full dependence on the
top quark mass is retained.  Such a calculation also allows a direct quantification of the breakdown
of the EFT approach.

Analytic results for the Higgs+3 parton amplitude in the full theory
have been known for a long time~\cite{Ellis:1987xu,Baur:1989cm}.  Corresponding results
for Higgs+4 parton amplitudes have been obtained in refs.~\cite{DelDuca:2001fn,Neumann:2016dny},
although in both cases expressions for at least some of the amplitudes were too long to report.
In addition there are several automatic procedures than can provide
numerical results for one-loop
amplitudes~\cite{Cullen:2014yla,Hirschi:2015iia,Greiner:2016awe,Buccioni:2019sur}.  The aim of this
paper is to present compact amplitudes for all contributing processes, 
\begin{eqnarray}
&& 0 \to gggg\Higgs \,, \\
&& 0 \to \qb q gg\Higgs \,, \\
&& 0 \to \qb q \qb^\prime q^\prime \Higgs \,, 
\end{eqnarray}
retaining all mass effects.  Compact analytic results for the $0 \to gggg\Higgs$
case when all the gluons have positive helicity have been published in
ref.~\cite{Ellis:2018hst}.  Although our result is therefore not new
{\it per se}, it is the first time that a compact publishable analytic
result has been obtained for all gluon helicities.  A calculation with
compact analytic formulae allows examination of the structure of the
amplitude for all values of the fermion mass.  It also has the
potential to lead to faster and more stable numerical evaluation of
the amplitude.  This would be a boon to calculations requiring this
amplitude in all regions of phase space, such as recent NLO
predictions for Higgs boson plus 1-jet production in the full
theory~\cite{Jones:2018hbb} and at large transverse momentum~\cite{Lindert:2018iug,Neumann:2018bsx}.
Although the results are quite compact, given the number of integral coefficients, this
paper is not easy to read. However we believe that it is
detailed enough that readers wishing to implement this amplitude in a
numerical program will find enough information to do so in our paper.

\section{Structure of the calculation}
\subsection{Definition of colour amplitudes}
The amplitude for the production of a Higgs boson and $n$ gluons can be expressed in 
colour-ordered sub-amplitudes as follows:
\begin{eqnarray}
        \label{exp}
        {\cal H}_n^{gggg}(\{p_i,h_i,c_i\})\,&=&\,i\frac{g_s^n}{16 \pi^2} \frac{m^2}{v}\sum_{\{1,2,\dots,n\}'}\;\tr\,(t^{c_1}t^{c_2} \dots t^{c_n})
        H_n^{\{c_i\}}(1^{h_1},2^{h_2},\ldots n^{h_n};\Higgs)\, ,
\end{eqnarray}
where the sum with the {\it prime}, $\sum_{\{1,2,\dots,n\}'}$, is over all
$(n-1)!$ {\em non-cyclic}  permutations of $1,2,\dots,n$ and the $t$ matrices are the SU(3) matrices in the fundamental representation normalized such that,
\begin{equation}    \label{normalization}
        \tr(t^a t^b)\;=\; \delta^{ab}.
\end{equation}
$m$ is the mass of the quark circulating in the loop.
Because of Bose symmetry it is sufficient to calculate one
permutation, and the other colour sub-amplitudes can be obtained by
exchange.

For the particular case at hand with four gluons Eq.~(\ref{exp}) becomes,
\begin{eqnarray}
        \label{explicitfor4}
        &&\cH^{gggg}_4(\{p_i,h_i,c_i\})\,=\,i\frac{g_s^4}{16 \pi^2} \bigg(\frac{m^2}{v}\bigg) 
\Bigg\{\big[ \tr\,(t^{c_1}t^{c_2}t^{c_3}t^{c_4})+ \tr\,(t^{c_1}t^{c_4}t^{c_3}t^{c_2}) \big] H_4^{1 2 3 4}(1^{h_1},2^{h_2},3^{h_3},4^{h_4};\Higgs) \nonumber \\
       & +&\big[ \tr\,(t^{c_1}t^{c_3}t^{c_4}t^{c_2})+ \tr\,(t^{c_1}t^{c_2}t^{c_4}t^{c_3}) \big] H_4^{1 3 4 2}(1^{h_1},2^{h_2},3^{h_3},4^{h_4};\Higgs) \nonumber \\
             &+&\big[ \tr\,(t^{c_1}t^{c_4}t^{c_2}t^{c_3})+ \tr\,(t^{c_1}t^{c_3}t^{c_2}t^{c_4}) \big] H_4^{1 4 2 3}(1^{h_1},2^{h_2},3^{h_3},4^{h_4};\Higgs) \Bigg\} \, .
\end{eqnarray}
Squaring the amplitude Eq.~(\ref{explicitfor4}) for a fixed helicity configuration and summing over colours we find
\begin{eqnarray}
\label{coloursum}
\sum_{\rm colours}\left|\cH^{gggg}_4\right|^2 &=& \bigg[\frac{g_s^4}{16 \pi^2} \bigg(\frac{m^2}{v}\bigg) \bigg]^2 (N^2-1) 
\Bigg\{2 N^2 \big(\left|H^{1234}_4\right|^2+\left|H^{1342}_4\right|^2+\left|H^{1423}_4\right|^2\big) \nonumber \\
                   &-&4 \frac{(N^2-3)}{N^2} \left|H^{1234}_4+H^{1342}_4+H^{1423}_4\right|^2 \Bigg\}\, ,
\end{eqnarray}
where $N$ is the dimensionality of the $SU(N)$ colour group, i.e.~$N=3$, and the labels for the helicity configuration
(as explicitly shown in Eq.~(\ref{explicitfor4})) have been suppressed.

The amplitude for the production of a Higgs boson, an antiquark, quark and two gluons is similarly decomposed into
colour-ordered amplitudes as follows,
\begin{eqnarray}
        \label{explicitfor2q}
{\cal H}^{{\qb}qgg}_4(\{p_i,h_i,c_i,j_i\})= i\, \frac{g_s^4}{16 \pi^2 } \Big(\frac{m^2}{v}\Big) && \Big[
  (t^{c_3}\,t^{c_4})_{j_2\,j_1} H^{34}_4(1^{h_1},2^{-h_1},3^{h_3},4^{h_4};\Higgs) \nonumber \\
&&+(t^{c_4}\,t^{c_3})_{j_2\,j_1} H^{43}_4(1^{h_1},2^{-h_1},3^{h_3},4^{h_4};\Higgs) \Big] \;.
\end{eqnarray}
In this paper we will give results for the colour-ordered amplitude $H_4^{34}$.  It is straightforward to obtain
$H_4^{43}$ from this through the parity operation (complex conjugation) and permutation of momentum labels.
Squaring and summing over colours yields,
\begin{equation}
\sum |{\cal H}^{{\qb}qgg}_4|^2 =\Big(\frac{g_s^4}{16 \pi^2 }\Big)^2 \Big(\frac{m^2}{v}\Big)^2
 \, (N^2-1) \, \left[ N \left( |H^{34}_{4}|^2 + |H^{43}_{4}|^2 \right)
  - \frac{1}{N} |H^{34}_{4}+H^{43}_{4}|^2 \right] \,,
\end{equation}
where the labelling of the helicity configuration shown in Eq.~(\ref{explicitfor2q}) has again been suppressed.

The four-quark amplitude takes the form,
\begin{equation}
{\cal H}^{4q}_4(\{p_i,h_i,j_i\})= i\, \frac{g_s^4}{16 \pi^2 } \Big(\frac{m^2}{v}\Big)\; (t^{c_1})_{j_2\,j_1}\;(t^{c_1})_{j_4\,j_3} 
H^{4q}_{4}(1^{h_1}_{{\qb}},2^{-h_1}_q,3^{h_3}_{{\qb}^\prime},4^{-h_3}_{q^\prime})
\end{equation}
where the helicities of the quarks are fixed by those of the antiquarks.Performing the sum over colours we then have,
\begin{equation}
\sum |{\cal H}^{4q}_4(h_1, h_3)|^2 =\Big(\frac{g_s^4}{16 \pi^2 }\Big)^2 \Big(\frac{m^2}{v}\Big)^2
 \, (N^2-1) \, |H^{4q}_{4}(h_1,h_3)|^2 
\end{equation}
for the case in which the quark lines have different flavours.  For the case of identical quarks we first introduce,
\begin{equation}
H^{4q^\prime}_{4}(h_1,h_3) = H^{4q}_{4}(1_{{\qb}}^{h_1},4_{q}^{-h_1},3_{{\qb}}^{h_3},2_{q}^{-h_3})
\end{equation}
The sum over the colours for the identical case is then,
\begin{eqnarray}
\sum |{\cal H}^{4q}_4|^2 &=& \Big(\frac{g_s^4}{16 \pi^2 }\Big)^2 \Big(\frac{m^2}{v}\Big)^2\, (N^2-1)\, \Bigg( 
|H^{4q}_{4}(h_1,h_3)|^2 +|H^{4q^\prime}_{4}(h_1,h_3)|^2 \nonumber \\
 && \qquad + \frac{\delta_{h_1h_3}}{N} \left(H^{4q}_{4}(h_1,h_3) H^{4q^\prime}_{4}(h_1,h_3)^*
                                            +H^{4q}_{4}(h_1,h_3)^* H^{4q^\prime}_{4}(h_1,h_3) \right) \Bigg)
\end{eqnarray}
where, as indicated, the term on the second line only contributes when the quarks have the same helicity.

\subsection{Decomposition into scalar integrals}
\label{Decomposition}
The colour-ordered sub-amplitudes can be expressed in terms of scalar integrals.  For the $0 \to gggg\Higgs$
sub-amplitude we have,
\begin{eqnarray} \label{fermionreduction}
H_4^{1234}(1^{h_1},2^{h_2},3^{h_3},4^{h_4};\Higgs) & = & \frac{\bar\mu^{4-n}}{r_\Gamma}\frac{1}{i \pi^{n/2}} \int {\rm d}^n \ell
 \, \frac{{\rm Num}(\ell)}{\prod_i d_i(\ell)} \nonumber \\
&=& \sum_{i,j,k,l} {e}_{i \x j \x k \x l}(1^{h_1},2^{h_2},3^{h_3},4^{h_4}) \, E_0(p_i,p_j,p_k,p_l;m) \nonumber \\
&+& \sum_{i,j,k} {d}_{i\x j\x k}(1^{h_1},2^{h_2},3^{h_3},4^{h_4}) \, D_0(p_i, p_j, p_k ;m)  \nonumber \\
&+& \sum_{i,j} {c}_{i\x j}(1^{h_1},2^{h_2},3^{h_3},4^{h_4}) \,  C_0(p_i,p_j ;m)   \nonumber \\
&+& \sum_{i} {b}_{i}(1^{h_1},2^{h_2},3^{h_3},4^{h_4}) \, B_0(p_i;m) + r(1^{h_1},2^{h_2},3^{h_3},4^{h_4})\, .
\end{eqnarray}
The scalar bubble ($B_0$), triangle ($C_0$), box ($D_0$) and pentagon ($E_0$) integrals, and the
constant $r_\Gamma$, are defined in Appendix~\ref{Integrals}.
$\bar{\mu}$ is an arbitrary mass scale, and $r$ are the rational terms.
The rank of a Feynman integral
is defined to be the number of powers of the loop momentum in the numerator. A scalar Feynman integral
has no powers of the loop momentum in the numerator, and is hence of rank zero.  All scalar
integrals are well known and readily evaluated using existing
libraries~\cite{Ellis:2007qk,vanHameren:2010cp,Carrazza:2016gav}.
The sums in the above equation scan over groupings of external gluons.
Thus, for example, the sum for the scalar triangle integrals will contain a term $c_{1\x234}$ which multiplies the 
scalar triangle integral $C_0(p_1,p_{234};m)$ where $p_{234}=p_2+p_3+p_4$.
The reduction in Eq.~(\ref{fermionreduction}) is written in $n$ dimensions, although at the end the amplitude is finite.
The individual bubble integrals contain ultra-violet singularities that are regulated using dimensional regularization.

In four dimensions the pentagon integral can be reduced to a sum of the five box integrals obtained by removing
one propagator~\cite{Melrose:1965kb,vanNeerven:1983vr,Bern:1993kr},
\begin{eqnarray} \label{pentagonreduction}
E_{0}(p_1,p_2,p_3,p_4;m)&=& 
 \mC_{1\times2\times3\times4}^{(1)}\,D_0(p_2,p_3,p_4;m)
+\mC_{1\times2\times3\times4}^{(2)}\,D_0(p_{12},p_3,p_4;m) \nonumber \\
&+&\mC_{1\times2\times3\times4}^{(3)}\,D_0(p_1,p_{23},p_4;m) \nonumber \\
&+&\mC_{1\times2\times3\times4}^{(4)}\,D_0(p_1,p_2,p_{34};m)
+\mC_{1\times2\times3\times4}^{(5)}\,D_0(p_1,p_2,p_3;m)\, .
\end{eqnarray}
Explicit forms for the pentagon reduction coefficients, $\mC_{1\times2\times3\times4}^{(i)}$, are
\begin{eqnarray}
\label{pentredcoeffs}
\mC_{1\times2\times3\times4}^{(1)}
&=&-\frac{1}{2}\,\frac{s_{23}\,s_{34}\,[2\,s_{12}\,s_{24}+s_{13}\,s_{24}+s_{34}\,s_{12}-s_{23}\,s_{14}]}{16\, \detS}\nonumber\\
\mC_{1\times2\times3\times4}^{(2)}
&=&-\frac{1}{2}\,\frac{s_{34}\,[s_{1234}\,s_{23}\,(s_{123}-2\,s_{12})+s_{123}\,(s_{34}\,(s_{123}-s_{23})+s_{12}\,(s_{234}+s_{23})-s_{234}\,s_{123})]}{16\, \detS}\nonumber\\
\mC_{1\times2\times3\times4}^{(3)}
&=&-\frac{1}{2}\,\frac{[s_{14}\,s_{23}-(s_{12}+s_{13})\,(s_{24}+s_{34})]\,[s_{34}\,s_{12}+s_{23}\,s_{14}-s_{13}\,s_{24}]}{16\, \detS}\nonumber\\
\mC_{1\times2\times3\times4}^{(4)}
&=&-\frac{1}{2}\,\frac{s_{12}\,[s_{1234}\,s_{23}\,(s_{234}-2\,s_{34})+s_{234}\,(s_{12}\,(s_{234}-s_{23})+s_{34}\,(s_{123}+s_{23})-s_{234}\,s_{123})]}{16\, \detS}\nonumber\\
\mC_{1\times2\times3\times4}^{(5)}
&=&-\frac{1}{2}\,\frac{s_{12}\,s_{23}\,[2\,s_{34}\,s_{13}+s_{13}\,s_{24}+s_{34}\,s_{12}-s_{23}\,s_{14}]}{16\, \detS}
\end{eqnarray}
The factor $\detS$ is the determinant of the matrix,
$\left[S_{1\x2\x3\x4}\right]_{ij}=[m^2-\frac{1}{2}(q_{i-1}-q_{j-1})^2]$,
where $q_i$ is the offset momentum, see Eq.~(\ref{denominators}).  It can be written as,
\begin{eqnarray}
\label{Sdef}
16\, \detS&=&s_{12}\,s_{23}\,s_{34}\,\big(s_{14}\,s_{23}-(s_{12}+s_{13})\,(s_{24}+s_{34})\big)+m^2\,\DeltaFour\,,\nonumber \\
\DeltaFour  &=& (s_{12}\,s_{34}-s_{13}\,s_{24}-s_{14}\,s_{23})^2-4\,s_{13}\,s_{14}\,s_{23}\,s_{24} \,.
\end{eqnarray}
As a result of Eq.~(\ref{pentagonreduction}), in four dimensions
the integral basis given in Eq.~(\ref{fermionreduction}) is overcomplete. The full amplitude can be described 
by the box, triangle and bubble integrals alone (+ rational terms). This is the specific choice made in this paper, but
the other choice to keep the redundant basis of Eq.~(\ref{fermionreduction}) is also perfectly viable.
In this paper we will work in a basis without pentagon integrals, but the box coefficients will in part display 
vestiges of their pentagon origin, through effective pentagon coefficients and the presence of the pentagon-to-box
reduction coefficients, $\mC_{1\times2\times3\times4}^{(i)}$.  This will be explained in detail in Section~\ref{pppm}.
Our decomposition of the sub-amplitudes is thus,
\begin{eqnarray} \label{fermionreductionnewsv}
H_4(1^{h_1},2^{h_2},3^{h_3},4^{h_4};\Higgs) 
&=& \sum_{i,j,k} {d}_{i\x j\x k}(1^{h_1},2^{h_2},3^{h_3},4^{h_4}) \, D_0(p_i, p_j, p_k ;m)  \nonumber \\
&+& \sum_{i,j} {c}_{i\x j}(1^{h_1},2^{h_2},3^{h_3},4^{h_4}) \,  C_0(p_i,p_j ;m)   \nonumber \\
&+& \sum_{i} {b}_{i}(1^{h_1},2^{h_2},3^{h_3},4^{h_4}) \, B_0(p_i;m) + r(1^{h_1},2^{h_2},3^{h_3},4^{h_4})\, .
\end{eqnarray}
which also applies for the $0 \to \qb q g g \Higgs$ sub-amplitude $H_4^{34}$ since it contains no 
pentagon diagrams in the first place.

\subsection{Unitarity methods}
\label{Unitarity_methods}
The modern treatment of one-loop amplitudes containing massive 
particles was pioneered 25 years ago in ref.~\cite{Bern:1995db}. 
Since that early paper a whole set of tools and methods have been
invented to deal with one-loop amplitudes (for an introduction and comprehensive
review see refs.~\cite{Dixon:2013uaa} and~\cite{Ellis:2011cr} respectively).
We shall apply many of them
in order to arrive at the simplest form for the Higgs + 4 parton amplitude.
This paper will present compact expressions for the coefficients in the four dimensional 
version of Eq.~(\ref{fermionreduction}) where the scalar pentagon integral has been expressed  
as a sum of box integrals.
The coefficients will be expressed in terms of spinor products. Our notation for spinor products is reported in 
Appendix~\ref{spinorsection}.

As we shall see below, there is an intimate connection between the
full one-loop calculation with a massive fermion and a suitably
normalized calculation performed with the Higgs boson coupling to four
partons via a loop of colour-triplet, massive scalar particles.  The
latter calculation with scalar intermediaries has two advantages.
First, the scalar calculation is completely free of Dirac algebra,
allowing more compact expressions to be maintained throughout the
calculation. This is useful if one can show the identity of the
coefficients of scalar integrals between the scalar and the
fermionic theories.  Second, the scalar calculation, unlike the
fermionic calculation, can be performed in the $m \to 0$ limit.  If it
can be shown that:
\begin{enumerate}
\item the result for a particular coefficient in the scalar theory
is identical to the result in the fermionic theory,
\item that particular coefficient is also independent of the mass,
\end{enumerate}
the value of the $m \to 0$ limit is established.

In order to perform the reduction to scalar integrals indicated in Eq.~(\ref{fermionreduction})
we use unitarity techniques to isolate the contribution of boxes~\cite{Britto:2004nc},
triangles~\cite{Forde:2007mi} and bubbles~\cite{Mastrolia:2009dr,Kilgore:2007qr,Davies:2011vt}.
Since bubble integrals satisfy both criteria enumerated above, their coefficients
are most easily calculated using a massless internal scalar loop.

Integrals that do not give rise to rational terms are said to be cut-constructible.
In general, $x$-point integrals are cut constructible in four dimensions if the
rank $r$ satisfies $r < {\rm max}[(x-1,2)]$.  Thus rank-3 pentagons,
rank-2 boxes, rank-1 triangles and bubbles are cut constructible.
Integrals that are not cut-constructible give rise to the rational
terms ($r$) in Eq.~(\ref{fermionreduction}).  In our case the rational terms can instead be
obtained by using already-computed results for the mass-dependent
coefficients of triangle integrals~\cite{Badger:2008cm}.

\subsection{Simplification techniques}
We now briefly describe two further techniques that are useful to help simplify the results
obtained using unitarity methods. Both methods exploit the fact that 
the kinematics of our process can be expressed as massless 6-point kinematics by decomposing 
the momentum of the Higgs boson into two light-like momenta, which for definiteness we call $p_5$,~$p_6$. 

Reduction of the analytic forms to simpler expressions is aided by the use of
momentum twistors~\cite{Hodges:2009hk,Badger:2013gxa,Badger:2016uuq,Hartanto:2019uvl}.
In this formalism each particle is described by a 4-component momentum twistor $Z(\lambda,\mu)$,
where $\lambda$ is the usual two-component holomorphic Weyl spinor (with
$\spa i.j  = \lambda_\alpha \lambda^\alpha$) and $\mu$ is a two-component object related
to dual momentum coordinates~\cite{Hodges:2009hk}.  Anti-holomorphic spinors ($\tilde\lambda_i$,
with $\spb i.j  = \tilde\lambda^{\dot{\alpha}} \tilde\lambda_{\dot{\alpha}}$) 
are obtained from these via the identity,
\begin{equation}
\label{antihol}
\tilde\lambda_i = \frac{\spa{i}.{(i+1)} \mu_{i-1} + \spa{(i+1)}.{(i-1)}\mu_i + \spa{(i-1)}.i \mu_{i+1}}
 {\spa{i}.{(i+1)} \spa{(i-1)}.{i} } \,.
\end{equation}
To describe an $n$-particle scattering amplitude there are thus $4n$ momentum twistor
components, of which only $3n-10$ are independent due to a $U(1)$ symmetry for each particle
and overall Poincar\'{e} symmetry.  We thus need 8 momentum-twistor variables
($x_1 \ldots x_8$) to describe our 6-point kinematics, which we choose to parametrize as,
\begin{equation}
Z =
\begin{pmatrix}
\lambda_1 & \lambda_2 & \lambda_3 & \lambda_4 & \lambda_5 & \lambda_6 \\
\mu_1 & \mu_2 & \mu_3 & \mu_4 & \mu_5 & \mu_6 
\end{pmatrix}
= 
\begin{pmatrix}
1~~ & 0~~ & y_1 ~& y_2     ~& y_3                                  ~& y_4 \\
0~~ & 1~~ & 1   ~& 1       ~& 1                                    ~& 1 \\
0~~ & 0~~ & 0   ~& x_5 x_6 ~& x_6                                  ~& 1 \\
0~~ & 0~~ & 1   ~& 1       ~& 1-\left(\frac{1-x_7}{x_2 x_5}\right) ~& -\left(\frac{1+x_8}{x_2 x_5 x_6}\right)
\end{pmatrix}
\,,
\end{equation}
where $y_i = \sum_{j=1}^i \Pi_{k=1}^j 1/x_k$.  The spinors involving our four massless partons are then given by, 
\begin{align}
&\spa1.2=-1,\, \spa1.3=-1,\, \spa1.4=-1,\, \spa2.3=1/x_1,\, \spa2.4=\frac{1+x_2}{x_1 x_2},\, \spa3.4=\frac{1}{x_1 x_2} \\
& \spb1.2=x_1,\, \spb1.3=x_1 x_8,\, \spb1.4=-\frac{x_1 (x_7 + x_8)}{x_5},\, \spb2.3=-x_1^2 x_2 x_5 x_6 ,\, 
\spb2.4=x_1^2 x_2 x_6,\, \spb3.4=-x_1^2 x_2 x_6 x_7 \,, \nonumber
\end{align}
where the $\spb i.j$ spinors in the second line have been derived with the aid of Eq.~(\ref{antihol}).
Note that the variables $x_3$ and $x_4$ are not present, leaving us with a rational parametrization of
our amplitude in terms of only 6 parameters.
Inverting we have
\begin{eqnarray}
x_1 &=& -\spa1.2\spb1.2,\, 
x_2 = +\frac{\spa2.3 \spa1.4}{\spa1.2 \spa3.4},\, 
x_3 = +\frac{\spa3.4 \spa1.5}{\spa1.3 \spa4.5},\, 
x_4 = -\frac{\spa4.5 \spa1.6}{\spa1.4 \spa5.6}, \nonumber \\
x_5 &=& -\frac{\spa1.3\spb2.3}{\spa1.4\spb2.4},\, 
x_6 = -\frac{\spa3.4 \spb2.4}{\spa1.3 \spb1.2},\,
x_7 = -\frac{\spa1.3 \spb3.4}{\spa1.2\spb2.4},\,
x_8 = +\frac{\spa1.3 \spb1.3}{\spa1.2\spb1.2}
\end{eqnarray}
where we see explicitly that the variables $x_3$ and $x_4$ involve momenta $p_5$ and $p_6$ and effectively
decouple in our case~\cite{Hartanto:2019uvl}.  In order to use this parametrization we first need to remove 
the overall phase of the coefficient corresponding to the helicities of the external
gluons, for example by multiplying by $\spa1.2^2 \spa3.4^2$ for the all-plus amplitude. 
The advantage of this approach is that the amplitude can now be simplified using straightforward algebra,
without needing to account for momentum conservation and Schouten identities to manipulate spinor strings. 
In this way overall factors can easily be identified and the true denominator structure of the coefficients
established.

The second method we adopt is to use high precision floating-point arithmetic to simplify our
analytic expressions~\cite{DeLaurentis:2019phz}. The study of singular and doubly singular limits in
complex phase space allows us to explore the singularity structure of the integral coefficients.
The integral coefficients can then be reconstructed by solving linear systems for the rational numerical coefficients of
generic spinor trial functions.  The helicities of the gluons impose constraints on the structure of the trial functions.
This method is particularly useful when unitarity techniques result in
lengthy expressions that are hard to treat using twistor variables, such as in the case of some
triangle and bubble coefficients. It is also useful to bypass the algebra involved in removing
artefacts of loop-momentum parametrizations, such as square roots and massless projections
of non-lightlike external momenta~\cite{Forde:2007mi, Ossola:2006us}.
This paper is the first instance where this method has been applied in the presence of massive
particles.

\section{Higgs boson production mediated by a coloured scalar}
\label{sec:scalar}
Consider a complex scalar field $\phi$ in the triplet representation 
of colour SU(3) coupled to a gluon field and to the Higgs boson, $\Higgs$.
The part of the QCD Lagrangian involving the field $\phi$ is
\begin{equation} \label{ScalarLagrangian}
{\cal L} = (D_\mu \phi^\dagger)_i (D^\mu \phi)_i -\lambda \phi^\dagger_i \phi_i \Higgs
\end{equation}
where $(D_\mu \phi)_i=(\partial_\mu\delta_{ij} + i g (t \cdot {\cal A}_\mu)_{ij}) \phi_j$.
The partial correspondence with the fermion theory emerges when setting $\lambda = -4 m^2/v$.  

We will calculate colour-ordered sub-amplitudes for the production of a Higgs boson and four gluons mediated by a scalar 
loop.  For the Higgs + 4 gluon amplitude this is,
\begin{eqnarray}
        \label{expbis}
        &&\cA^{gggg}_4(\{p_i,h_i,c_i\})\,=\,i\frac{g_s^4}{16 \pi^2} \left(-\frac{\lambda}{4}\right) 
\Bigg\{\big[ \tr\,(t^{c_1}t^{c_2}t^{c_3}t^{c_4})+ \tr\,(t^{c_1}t^{c_4}t^{c_3}t^{c_2}) \big] A_4^{1 2 3 4}(1^{h_1},2^{h_2},3^{h_3},4^{h_4};\Higgs) \nonumber \\
       &+&\big[ \tr\,(t^{c_1}t^{c_3}t^{c_4}t^{c_2})+ \tr\,(t^{c_1}t^{c_2}t^{c_4}t^{c_3}) \big] A_4^{1 3 4 2}(1^{h_1},2^{h_2},3^{h_3},4^{h_4};\Higgs) \nonumber \\
       &+&\big[ \tr\,(t^{c_1}t^{c_4}t^{c_2}t^{c_3})+ \tr\,(t^{c_1}t^{c_3}t^{c_2}t^{c_4}) \big] A_4^{1 4 2 3}(1^{h_1},2^{h_2},3^{h_3},4^{h_4};\Higgs) \Bigg\}
\end{eqnarray}
and the colour amplitudes have the decomposition,
\begin{eqnarray} \label{scalarreduction}
A_4^{1234}(1^{h_1},2^{h_2},3^{h_3},4^{h_4};\Higgs) & = & \frac{\bar\mu^{4-n}}{r_\Gamma}\frac{1}{i \pi^{n/2}} \int {\rm d}^n \ell
 \, \frac{{\rm Num}(\ell)}{\prod_i d_i(\ell)} \nonumber \\
&=& \sum_{i,j,k,l} \tilde{e}_{i \x j \x k \x l}(1^{h_1},2^{h_2},3^{h_3},4^{h_4}) \, E_0(p_i,p_j,p_k,p_l;m) \nonumber \\
&+& \sum_{i,j,k} \tilde{d}_{i\x j\x k}(1^{h_1},2^{h_2},3^{h_3},4^{h_4}) \, D_0(p_i, p_j, p_k ;m)  \nonumber \\
&+& \sum_{i,j} \tilde{c}_{i\x j}(1^{h_1},2^{h_2},3^{h_3},4^{h_4}) \,  C_0(p_i,p_j ;m)   \nonumber \\
&+& \sum_{i} \tilde{b}_{i}(1^{h_1},2^{h_2},3^{h_3},4^{h_4}) \, B_0(p_i;m) + \tilde{r}(1^{h_1},2^{h_2},3^{h_3},4^{h_4})\, .
\end{eqnarray}
Thus the tilde indicates that we are referring to an amplitude mediated by a scalar field.
For the case of the triangle coefficients we divide the coefficient 
into two pieces, to separate the mass dependence. Thus for both the fermion- and scalar-mediated loops we have, 
\begin{eqnarray} \label{c0andc2}
c_{i\x j}&=&c_{i \x j}^{(0)}+m^2 \, c_{i \x j}^{(2)} \\
\tilde{c}_{i\x j}&=&\tilde{c}_{i \x j}^{(0)}+m^2 \, \tilde{c}_{i \x j}^{(2)}
\label{c0andc2scalar}
\end{eqnarray}
In Eq.~(\ref{expbis}) we have chosen the normalization so that in some cases the coefficients $c_{i\x j}=\tilde{c}_{i\x j}$
and in addition in all cases $c_{i \x j}^{(2)}=\tilde{c}_{i \x j}^{(2)}$, $b_{i}=\tilde{b}_i$ and $r=\tilde{r}$. 
Thus we can perform certain parts of the calculations in the scalar theory. We further have that $\tilde{b}_i$ are independent of the mass $m$,
so that they can be calculated in the massless scalar theory. In addition, for the case at hand the rational terms are fully fixed by
$\tilde{c}_{i \x j}^{(2)}$,
\begin{equation}
\label{rattrick}
r(1^{h_1},2^{h_2},3^{h_3},4^{h_4})= \frac{1}{2} \sum_{i,j} \tilde{c}_{i\x j}^{(2)}(1^{h_1},2^{h_2},3^{h_3},4^{h_4})\, ,
\end{equation}
where the sum runs over all non-zero $\tilde{c}^{(2)}_{i\x j}$ for a particular helicity.

\subsection{Relationship of the fermion theory to the scalar theory} 
In order to elucidate the relationship between the fermion and scalar theories~\cite{Bern:1994cg} it is instructive
to review the steps previously used to demonstrate their similarity using the second order formalism~\cite{Morgan:1995te}.
Following ref.~\cite{Morgan:1995te} we define the quantity ${\bf A}$ to represent the combination of 
the numerator part of a fermion propagator with momentum $\ell$ and a gluon-quark-antiquark vertex at
which momentum $p_1$ flows out along the gluon line
\begin{equation} \label{Morgan}
{\bf A}^{\mu}=(\slsh{\ell}+m) \gamma^\mu 
  = (\slsh{\ell}+\frac{1}{2} \slsh{p}_1-\frac{1}{2} \slsh{p}_1+m) \gamma^\mu =(2 \ell^\mu+p_1^\mu){\bf 1} 
 - \frac{1}{2}[\slsh{p}_1,\gamma^\mu] -\gamma^\mu (\slsh{\ell}+\slsh{p}_1-m)   \, .
\end{equation}
\begin{figure}[t]
\begin{center}
\includegraphics[angle=270,width=0.8\textwidth]{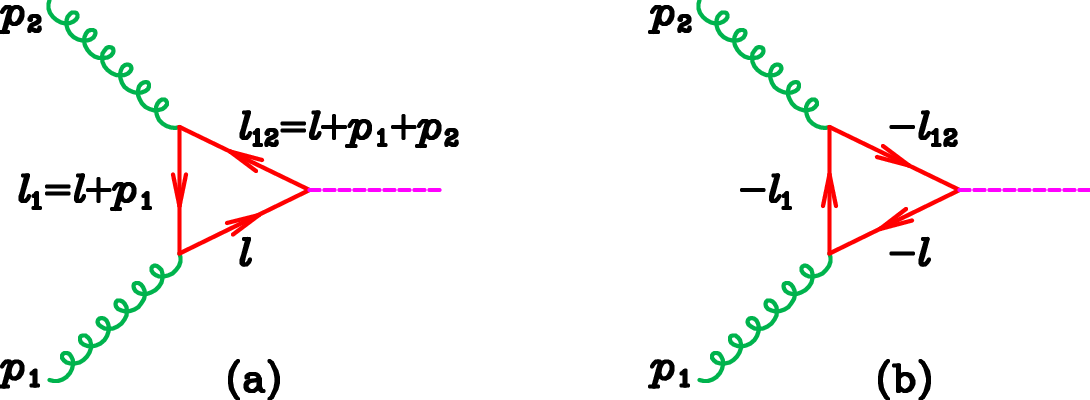}
\caption{Triangle diagrams showing the production of a Higgs boson by gluon fusion}
\label{triangle}
\end{center}
\end{figure}
The first term on the right hand side of Eq.~(\ref{Morgan}) already resembles the vertex 
for a gluon (on-shell or off-shell) interacting with a scalar field.
Now consider the integrand of a 
triangle diagram for a Higgs boson coupled to two gluons
as shown in Fig.~(\ref{triangle}a),
\begin{equation}
\label{triangle1}
(-1) \times t^{c_1} t^{c_2}\times 
\frac{1}{m} \frac{\tr \{ (\slsh{\ell}+m)\gamma^{\mu_1}(\slsh{\ell}_1+m)\gamma^{\mu_2}(\slsh{\ell}_{12}+m) \}}{D(\ell) \, D(\ell_1)\, D(\ell_{12})}
 \nonumber \\
\end{equation}
where our notation for the loop momenta and propagator factors $D(\ell)$ is given in
Appendix~\ref{Integrals}.
The minus sign is included because of the fermion loop and an overall factor of $1/m$ has been
included for convenience.  By using the decomposition in Eq.~(\ref{Morgan}) twice, the
expression in Eq.~(\ref{triangle1}) can be brought into the form,
\begin{equation}
(-1) \times t^{c_1} t^{c_2}\times \Bigg[\frac{ \tr \{ {\bf B}^{\mu_1}(\ell,\ell_1) {\bf B}^{\mu_2}(\ell_{1},\ell_{12})\}}{D(\ell) \, D(\ell_1)\, D(\ell_{12})}
          -\frac{\tr\{ \gamma^{\mu_1}\gamma^{\mu_2}\}}{D(\ell) \, D(\ell_{12})} \Bigg]  \, ,
\end{equation}
where ${\bf B}$ is a four-by-four matrix-valued function.  It contains a convection term
(as expected for a scalar field) and a spin term,
\begin{eqnarray}
\label{convectionplusspin}
{\bf B}^\mu(\ell,\ell_1)&=&(\ell^\mu+\ell_1^\mu) {\bf 1}  +\frac{1}{2} [(\slsh{\ell}-\slsh{\ell_1}), \gamma^\mu] \equiv
           (\ell^\mu+\ell_1^\mu)  {\bf 1}  -\frac{1}{2} [(\slsh{p}_1, \gamma^\mu]  \, .
\end{eqnarray}
Note that the spin term is independent of the loop momentum.
The result for the companion triangle as shown in Fig.~(\ref{triangle}b) is
\begin{eqnarray}
&&(-1) \times t^{c_2} t^{c_1}
\times \frac{1}{m} \frac{\tr\{ (-\slsh{\ell_{12}}+m)\gamma^{\mu_2}(-\slsh{\ell}_1+m)\gamma^{\mu_1}(-\slsh{\ell}+m) \}}{D(\ell) \, D({\ell_1})\, D({\ell_{12})}}
 \nonumber \\
&=&(-1) \times t^{c_2} t^{c_1}\times \Big[\frac{\tr\{ {\bf B}^{\mu_2}(-\ell_{12},-\ell_1) {\bf B}^{\mu_1}(-\ell_{1},-\ell)\}}{D(\ell) \, D(\ell_1)\, D(\ell_{12})}
          -\frac{\tr\{ \gamma^{\mu_2}\gamma^{\mu_1}\}}{D(\ell) \, D(\ell_{12})}\Big] \, .
\end{eqnarray}
Adding both diagrams, dropping vanishing terms and exploiting the cyclicity of the trace, the
integrand appearing in the full amplitude is,  
\begin{eqnarray} \label{trianglefull}
 (-1) \Bigg\{
&&t^{c_1} t^{c_2}\times \Bigg[ \frac{(\ell^{\mu_1}+\ell_1^{\mu_1})(\ell_1^{\mu_2}+\ell_{12}^{\mu_2}) \tr \{{\bf 1} \}
+\frac{1}{4} \tr \{[\slsh{p}_1,\gamma^{\mu_1}] [\slsh{p}_2,\gamma^{\mu_2}]\}}
{D(\ell) \, D(\ell_1)\, D(\ell_{12})}\Bigg]\nonumber \\
&+&t^{c_2} t^{c_1} \times \Bigg[ \frac{(\ell^{\mu_1}+\ell_1^{\mu_1})(\ell_1^{\mu_2}+\ell_{12}^{\mu_2}) \tr \{{\bf 1} \}
+\frac{1}{4} \tr \{[\slsh{p}_1,\gamma^{\mu_1}] [\slsh{p}_2,\gamma^{\mu_2}]\}}
{D(\ell) \, D(\ell_1)\, D(\ell_{12})}\Bigg] \nonumber \\
 &-& (t^{c_1} t^{c_2}+t^{c_2} t^{c_1}) g^{\mu_1 \mu_2}\frac{{\tr\{\bf 1}\}}{D(\ell) \, D(\ell_{12})} \Bigg\} \, .
\end{eqnarray}
If we drop the terms involving the commutators of gamma matrices, the 
fermionic loop (after removing an overall factor of $m$) 
can be written as the effect of a (suitably normalized) scalar triangle, with 3-point and 4-point (seagull) vertices. 
The full fermionic theory requires the inclusion of the additional spin flip terms, given by commutators. These additional terms 
do not involve the loop momentum and are thus of lower rank. Note also that there is no 
explicit mass dependence in Eq.~(\ref{trianglefull}). Dependence on the mass $m$ will be generated by the reduction to scalar integrals.
In contrast to the full fermionic theory, we may also consider the scalar theory in the massless case.

Iterating this argument for a larger number of gluons, it can be shown that separation of the full
fermionic theory into  a suitably normalized scalar theory, plus spin terms involving gamma matrix
commutators of rank lower by two powers of $\ell$, continues to hold.
The scalar theory is obtained by dropping all of the spin terms, c.f. Eq.~(\ref{convectionplusspin}).
Thus the full amplitude can be written as the sum of the scalar theory and a correction of lower rank, $\Delta F $,
\begin{equation}
{\rm Fermion~theory} = {\rm Scalar~theory} + \Delta F
\end{equation}

The difference between the scalar theory and the full fermionic theory $\Delta F$ 
is of lower rank. Although the scalar theory contains rank-4 pentagons, rank-3 boxes, and rank-2 triangles, 
$\Delta F$ contains only rank-2 pentagons, rank-1 boxes and rank-0 triangles. 
This has several important consequences:
\begin{enumerate}
\item $\Delta F$ is cut constructible. 
\item $\Delta F$ gives no contribution to bubble integral coefficients. Bubble integral coefficients
can thus be calculated in the scalar theory.
\item $\Delta F$ gives no contribution to the $m^2$ contributions to triangle coefficients, $c^{(2)}_{i\x j}$.
\item $\Delta F$ gives no contribution to certain $m^0$ triangle coefficients, $c^{(0)}_{i\x j}$,
\end{enumerate}

Exploiting these facts provides the following identities for the $gggg\Higgs$ case.
\begin{eqnarray}
c_{3\x4}(1^+,2^+,3^+,4^-) &=&   \tilde{c}_{3\x4}(1^+,2^+,3^+,4^-)  \, , \nonumber \\
c_{2\x34}(1^+,2^+,3^+,4^-) &=&  \tilde{c}_{2\x34}(1^+,2^+,3^+,4^-)  \, , \nonumber \\
c_{1\x43}(1^+,2^+,3^+,4^-) &=&  \tilde{c}_{1\x43}(1^+,2^+,3^+,4^-)  \, , 
\end{eqnarray}
\begin{eqnarray}
c_{3\x4}(1^+,2^-,3^+,4^-) &=&   \tilde{c}_{3\x4}(1^+,2^-,3^+,4^-) \, ,  \nonumber \\
c_{2\x34}(1^+,2^-,3^+,4^-) &=&  \tilde{c}_{2\x34}(1^+,2^-,3^+,4^-) \, , 
\end{eqnarray}
\begin{eqnarray}
c_{2\x3}(1^+,2^+,3^-,4^-) &=&   \tilde{c}_{2\x3}(1^+,2^+,3^-,4^-)  \, , \nonumber \\
c_{1\x23}(1^+,2^+,3^-,4^-) &=&  \tilde{c}_{1\x23}(1^+,2^+,3^-,4^-) \, , 
\end{eqnarray}
i.e. all triangles that do not have the Higgs boson as an external leg can be calculated fully in the 
scalar theory. In appendix~\ref{scalarsection} we reproduce several results for tree graphs involving 
massive scalars, Higgs bosons, gluons and a quark-antiquark pair. These results are useful in applying
unitarity to calculate loop diagrams.  

\section{Coefficients for $H^{1234}_4(g^+,g^+,g^+,g^+;\Higgs)$}
\label{pppp}

The reduction of a scalar pentagon integral into a sum of box integrals shown in Eq.~(\ref{pentagonreduction})
applies in four dimensions.  Therefore an extraction of the scalar pentagon integral coefficient
must be performed by making use of unitarity methods in $d$ dimensions.  To this end the loop momentum
is expressed most generally as,
\begin{equation}
\label{loopdecomposition}
\ell^\nu = \alpha p_1^\nu + \beta p_2^\nu + \frac{\gamma}{2}\spab1.\gamma^\nu.2 + \frac{\delta}{2}\spab2.\gamma^\nu.1
 + \ell_\epsilon^\nu
\end{equation}
where $\ell_\epsilon$ represents the excursion beyond four dimensions, with $\ell_\epsilon^2 = -\mu^2$.
Putting the propagators on shell determines $\alpha$, $\beta$,  $\gamma$, $\delta$ and $\mu^2$.
Parametrizing the amplitude at hand with the decomposition in Eq.~(\ref{loopdecomposition}) we thus find
the pentagon coefficient,
\begin{equation}
\label{epppp}
e_{1\x2\x3\x4} (1^+,2^+,3^+,4^+) = \frac{(m^2+\mu^2)\,\left(s_{1234}-4 (m^2 + \mu^2)\right) \, \TrfourgR1.2.3.4 }{\spa1.2\spa2.3\spa3.4\spa4.1} 
\end{equation}
where the trace functions are defined by,
\begin{eqnarray} \label{tracenotation}
{\rm tr}_{+}\{1\, 2\, 3\, 4\} &=& {\rm tr} \{\gamma_R \,\sp_1\,\sp_2,\sp_3,\sp_4 \} 
 = \spb1.2\,\spa2.3 \, \spb3.4 \spa4.1 \,,\nonumber \\
{\rm tr}_{-}\{1\, 2\, 3\, 4\} &=& {\rm tr} \{\gamma_L \,\sp_1\,\sp_2,\sp_3,\sp_4 \} 
 = \spa1.2\,\spb2.3 \, \spa3.4 \spb4.1 \,,
\end{eqnarray}
and $\gamma_{R/L}=(1\pm\gamma_5)/2$. The identities on the far right of Eq.~(\ref{tracenotation}) holds only for lightlike $p_i$.
The value of $\mu^2$ is fixed by the constraint $\ell^2 = m^2$ 
(implying $m^2 + \mu^2 = -\gamma\delta \, s_{12}$) but we have left the coefficient in Eq.~(\ref{epppp})
in this form in order to emphasise the $d$-dimensional nature of the result.
We choose to write our amplitude in terms of an effective pentagon coefficient $\ehat$ that corresponds to
the four-dimensional limit $\mu^2 \to 0$,
\begin{equation}
\ehat_{1\x2\x3\x4} (1^+,2^+,3^+,4^+)=   \frac{m^2\,(s_{1234}-4 m^2) \TrfourgR1.2.3.4 }{\spa1.2\spa2.3\spa3.4\spa4.1} 
\end{equation}
This leads to a very compact form for the complete amplitude~\cite{Ellis:2018hst},
\begin{eqnarray} \label{Higgs4gluons1}
H_4^{1234} (1^+,2^+,3^+,4^+;\Higgs)&=&  \Bigg\{ \frac{4 m^2-s_{1234}}{\spa1.2\spa2.3\spa3.4\spa4.1} 
 \Big[- \TrfourgR1.2.3.4  m^2 E_0(p_1,p_2,p_3,p_4;m)   \nonumber \\
&+& \frac{1}{2} ((s_{12}+s_{13})(s_{24}+s_{34})-s_{14}s_{23}) 
  D_0(p_1,p_{23},p_4;m)\nonumber \\
&+&\frac{1}{2} s_{12} s_{23} D_0(p_{1},p_{2},p_{3};m)\nonumber \\
&+& (s_{12}+s_{13}+s_{14}) C_0(p_1,p_{234};m)\Big]
 +2 \frac{s_{12}+s_{13}+s_{14}}{\spa1.2\spa2.3\spa3.4\spa4.1}  
\Bigg\}\nonumber \\
&+&\Bigg\{ 3~{\rm cyclic~permutations}\Bigg\} \, .
\end{eqnarray}
Although this expression, that includes the scalar pentagon integral, is very simple, we do not follow this
approach for the other helicity choices in the following sections.  Instead, since in the end our aim is to
produce a numerical code to calculate this amplitude, we feel that the structure is more straight-forward
working only in terms of boxes, triangles and bubble coefficients.
Adopting this approach also for this amplitude means that the minimal set of
coefficients that must be specified corresponds to the ones shown in
the first and third columns of Table~\ref{table:pppp}.   Note that coefficients of integrals that could in
principle appear but that are not specified in this table (and in subsequent tables in later sections) vanish.
There is some ambiguity in the naming convention for the integral coefficients.
For a given colour-ordered amplitude the external legs will appear in
cyclic or anti-cyclic order, c.f.~Eq.~(\ref{explicitfor4}). 
For example, $d_{1\x2\x34}$ in Table~\ref{table:pppp} could equally well
be written as $d_{43\x2\x1}$. Our convention is that the vertex
containing more than one gluon, should appear last in the name of the
coefficient. Where the compound vertex is in the centre, we have chosen the cyclic ordering.

\begin{table}
\begin{center}
\begin{tabular}{|l|l||l|l|}
\hline
Coefficient          & Related coefficients &Coefficient          & Related coefficients\\
\hline
  $d_{1\x2\x34}$ & $d_{2\x3\x41},d_{3\x4\x12},d_{4\x1\x23},$   &   $c_{1\x234}$   & $c_{2\x341},c_{3\x412},c_{4\x123}$\\
                 & $d_{1\x4\x32},d_{2\x1\x43},d_{3\x2\x14},d_{4\x3\x21}$ &&\\
  $d_{1\x23\x4}$ & $d_{2\x34\x1},d_{3\x41\x2},d_{4\x12\x3}$ &&\\
  $d_{1\x2\x3}$  & $d_{2\x3\x4},d_{3\x4\x1},d_{4\x1\x2}$ && \\
\hline
\end{tabular}
\caption{Minimal set of integral coefficients for $1_g^+\,2_g^+\,3_g^+\,4_g^+$.}
\label{table:pppp}
\end{center}
\end{table}

\subsection{Boxes} 
\subsubsection{$d_{1\x2\x34}$}
\begin{eqnarray}
  d_{1\x2\x34}(1^+,2^+,3^+,4^+) &=& \mC_{1\x2\x3\x4}^{(4)} \, \ehat_{\{1^+\x2^+\x3^+\x4^+\}}
\end{eqnarray}
\subsubsection{$d_{1\x23\x4}$}
\begin{eqnarray}
  d_{1\x23\x4}(1^+,2^+,3^+,4^+) &=& \mC_{1\x2\x3\x4}^{(3)}\,\ehat_{\{1^+\x2^+\x3^+\x4^+\}} \nonumber \\
                                &+&\frac{1}{2} \frac{(4\,m^2-s_{1234})}{\spa1.2\spa2.3\spa3.4\spa4.1}\,[(s_{12}+s_{13})\,(s_{24}+s_{34})-s_{14}\,s_{23}]
\end{eqnarray}
\subsubsection{$d_{1\x2\x3}$}
\begin{eqnarray}
d_{1\x2\x3}(1^+,2^+,3^+,4^+)  &=& \mC_{4\x1\x2\x3}^{(1)}\,\ehat_{\{4^+\x1^+\x2^+\x3^+\}} + \mC_{1\x2\x3\x4}^{(5)}\,\ehat_{\{1^+\x2^+\x3^+\x4^+\}}\nonumber  \\
                              &+& \frac{1}{2}  \frac{(4\,m^2-s_{1234})}{\spa1.2\spa2.3\spa3.4\spa4.1}\,s_{12}\,s_{23} 
\end{eqnarray}

\subsection{Triangles} 
\subsubsection{$c_{1\x234}^{(0)},c_{1\x234}^{(2)}$}
\begin{eqnarray}
c_{1\x234}^{(0)}(1^+,2^+,3^+,4^+)&=&-(s_{12}+s_{13}+s_{14}) \frac{s_{1234}}{\spa1.2\,\spa2.3\,\spa3.4\,\spa4.1}
\end{eqnarray}
\begin{eqnarray}
c_{1\x234}^{(2)}(1^+,2^+,3^+,4^+)&=&4 (s_{12}+s_{13}+s_{14}) \frac{1}{\spa1.2\,\spa2.3\,\spa3.4\,\spa4.1}
\end{eqnarray}
In presenting these results we have adopted the notation defined in Eq.~(\ref{c0andc2}) to separately quote the mass-independent
and mass-dependent parts.

\subsection{Rational terms} 
\begin{eqnarray}
      r(1^+,2^+,3^+,4^+) &=& \frac{1}{2}\,\big[c_{1\x234}^{(2)}(1^+,2^+,3^+,4^+)+c_{1\x234}^{(2)}(2^+,3^+,4^+,1^+)\nonumber \\
                                         && +c_{1\x234}^{(2)}(3^+,4^+,1^+,2^+)+c_{1\x234}^{(2)}(4^+,1^+,2^+,3^+)\big]\nonumber \\
                         &=& 4\,\frac{s_{1234}}{\spa1.2\spa2.3\spa3.4\spa4.1} 
\end{eqnarray}

\section{Coefficients for $H^{1234}_4(g^+,g^+,g^+,g^-;\Higgs)$}
\label{pppm}
Following the procedure outlined in Section~\ref{pppp} to obtain the pentagon coefficients yields, for 
this particular helicity combination, 
\begin{equation}
\label{pppmpent}
e_{\{1^+\x2^+\x3^+\x4^-\}} = - s_{12} s_{34} \left(s_{123}-4 (m^2+\mu^2)\right)
 \,\Bigg[\frac{\spb2.3 \, \spab4.(2+3).1}{\Trfourgfive1.2.3.4}\Bigg]^2 \, ,\\
\end{equation}
where we have introduced the notation,
\begin{equation}
\label{tr5def}
\Trfourgfive1.2.3.4 = {\rm tr} \{\gamma_5 \,\sp_1\,\sp_2\,\sp_3\,\sp_4 \}
 = \spb1.2\,\spa2.3\,\spb3.4\,\spa4.1 - \spa1.2\,\spb2.3\,\spa3.4\,\spb4.1\, .
\end{equation}
Note that the last identity, Eq.~(\ref{tr5def}), only applies for the case of lightlike momenta.
At this point we could follow the same strategy as in the previous section and take the limit $\mu^2 \to 0$
to obtain an effective pentagon coefficient $\ehat$.
However the appearance of the factor $\Trfourgfive1.2.3.4$ in the denominator of Eq.~(\ref{pppmpent})
is unpalatable since it is an
unphysical singularity.  In the amplitude its presence is compensated by corresponding factors in box
coefficients, and separating contributions in this way can lead to a loss of numerical precision.

As an alternative solution, we choose to modify the coefficient itself in such a way that
these factors are eliminated.   We do so by noting that the denominator can be expressed as,
\begin{equation}
\big(\Trfourgfive1.2.3.4\big)^2=(s_{12}\,s_{34}-s_{14}\,s_{23}-s_{13}\,s_{24})^2-4\,s_{14}\,s_{23}\,s_{24}\,s_{13}
 = \DeltaFour\, ,
\end{equation}
where $\DeltaFour$ has already been introduced in Eq.~(\ref{Sdef}). In fact, rearranging that equation and
making use of spinor notation we have,
\begin{equation}
\label{tr5removal}
s_{12} s_{23} s_{34} \spab1.(2+3).4\spab4.(2+3).1 =m^2 \, \big(\Trfourgfive1.2.3.4\big)^2 - 16\, \detS \, .
\end{equation}
We may use this equation to remove the denominator $\big(\Trfourgfive1.2.3.4\big)^2$ in part of
the pentagon coefficient in Eq.~(\ref{pppmpent}). We use Eq.~(\ref{tr5removal})
to eliminate one factor of $\spab4.(2+3).1$ from Eq.~(\ref{pppmpent}). This generates two terms,
one of which is free from the denominator $\big(\Trfourgfive1.2.3.4\big)^2$, which we identify as the
effective pentagon coeffcient.
The second term that is introduced contains a factor of $\detS$;  this neatly cancels the denominator factor
involved when reducing the pentagon integral to boxes (c.f. Eq.~({\ref{pentredcoeffs})) such that
factors of $1/\big(\Trfourgfive1.2.3.4\big)^2$ can be explicitly absorbed into the box coefficients and cancelled.

In this way we arrive at effective pentagon coefficients,
\begin{eqnarray}
\ehat_{\{1^+\x2^+\x3^+\x4^-\}} &=& (s_{123}-4 m^2)\, m^2 \,\Bigg[\frac{\spb2.3 \, \spab4.(2+3).1}{\spa2.3 \, \spab1.(2+3).4}\Bigg]\, , \\
\ehat_{\{4^-\x1^+\x2^+\x3^+\}} &=& \ehat_{\{1^+\x2^+\x3^+\x4^-\}}\{1\leftrightarrow3\}\, , \\
\ehat_{\{2^+\x3^+\x4^-\x1^+\}} &=& -m^2 \frac{\spb2.3}{\spa2.3\spb3.4}
 \Bigg( \frac{\spb2.3\spab2.(3+4).1\spab4.(1+3).2}{\spab1.(3+4).2}
	+\spb1.3\spab4.(2+3).1\nonumber \\
      &&+4m^2\frac{\spb2.3\spa3.4\spab2.(3+4).1}{\spa2.3\spab1.(3+4).2}\Bigg)\, ,\\
\ehat_{\{3^+\x4^-\x1^+\x2^+\}} &=& \ehat_{\{2^+\x3^+\x4^-\x1^+\}}\{1\leftrightarrow3\}\, .
\end{eqnarray}
We will see that the box coefficients take a particularly simple form when written this way.

The minimal set of remaining integral coefficients to determine this amplitude
is shown in the first and third columns of Table~\ref{table:pppm}.
The related coefficients included in the table are determined by using symmetry
properties of the amplitude and relabelling momenta.  These are mostly straightforward
except for the coefficient $b_{341}$ where we have found the relation,
\begin{equation}
b_{341}(1^+,2^+,3^+,4^-) = -b_{234}(2^+,3^+,1^+,4^-)-b_{234}(2^+,1^+,3^+,4^-) \,.
\end{equation}

\begin{table}
\begin{center}
\begin{tabular}{|l|l||l|l|}
\hline
Coefficient          & Related coefficients & Coefficient          & Related coefficients \\
\hline
  $d_{1\x2\x34}$ & $d_{3\x2\x14}$&                             $ c_{3\x4}$   & $c_{4\x1}$  \\  
  $d_{1\x4\x32}$ & $d_{3\x4\x12}$&                             $ c_{2\x34}$  & $c_{2\x14}$ \\  
  $d_{2\x1\x43}$ & $d_{2\x3\x41}$&                             $ c_{1\x43}$  & $c_{3\x41}$ \\  
  $d_{2\x34\x1}$ & $d_{3\x41\x2}$&                             $ c_{4\x123}$ & \\              
  $d_{4\x3\x21}$ & $d_{4\x1\x23}$&                             $ c_{1\x234}$ & $c_{3\x412}$ \\ 
  $d_{1\x23\x4}$ & $d_{4\x12\x3}$&                             $ c_{2\x341}$ & \\               
  $d_{2\x3\x4}$  & $d_{4\x1\x2}$ &                             $ c_{12\x34}$ & $c_{23\x41}$ \\  
  $d_{1\x2\x3}$  &               &                             $ b_{34}$     & $b_{14}$ \\            
  $d_{3\x4\x1}$  &               &                             $ b_{234}$    & $b_{412},b_{341}$\\ 
                 &               &                             $ b_{1234}$   & \\ 
\hline
\end{tabular}
\caption{Minimal set of integral coefficients for $1_g^+\,2_g^+\,3_g^+\,4_g^-$.}
\label{table:pppm}
\end{center}
\end{table}

\subsection{Boxes}
\subsubsection{$d_{1\x2\x34}$}
\begin{eqnarray}
  d_{1\x2\x34}(1^+,2^+,3^+,4^-) &=& \mC_{1\x2\x3\x4}^{(4)}\,\ehat_{\{1^+\x2^+\x3^+\x4^-\}} \nonumber \\
               &+&\frac{1}{2}\frac{\spb1.2\spb2.3\spab1.(2+4).3}{\spb3.4\spab1.(3+4).2\spab1.(2+3).4}\Bigg(\spb2.3s_{1234}-4\frac{\spab1.(2+4).3}{\spa1.2}m^2\Bigg)\nonumber \\
               &-&\frac{1}{2}\frac{\spb1.2\spa2.4\spab4.(2+3).1}{\spa2.3\spa3.4\spab2.(3+4).1}\Bigg(\spab4.(2+3).1+4\frac{\spa2.4}{\spa1.2}m^2\Bigg) 
\end{eqnarray}
\subsubsection{$d_{1\x4\x32}$}
\begin{eqnarray}
  d_{1\x4\x32}(1^+,2^+,3^+,4^-) &=& \mC_{2\x3\x4\x1}^{(2)}\,\ehat_{\{2^+\x3^+\x4^-\x1^+\}} \nonumber \\
                              &+&\frac{1}{2}\frac{\spb2.3s_{14}s_{234}}{\spa2.3^2\spb3.4\spab1.(3+4).2\spab1.(2+3).4}(4m^2s_{234}-s_{23}s_{1234})
\end{eqnarray}
\subsubsection{$d_{2\x1\x43}$}
\begin{eqnarray}
  d_{2\x1\x43}(1^+,2^+,3^+,4^-) &=& \mC_{3\x4\x1\x2}^{(2)}\,\ehat_{\{3^+\x4^-\x1^+\x2^+\}} \nonumber \\
                                &+&\frac{1}{2}\frac{\spb1.2\spab4.(1+3).2^2}{\spa1.2\spa3.4\spab1.(3+4).2\spab3.(1+4).2}(4m^2\spa1.4-\spa1.2\spab4.(1+3).2) \nonumber \\
                                &+&\frac{1}{2}\frac{\spb1.2\spb1.3^2}{\spa1.2\spb1.4\spb3.4\spab2.(3+4).1}(4m^2\spab2.(1+4).3-\spa2.1\spb1.3s_{1234})
\end{eqnarray}
\subsubsection{$d_{2\x34\x1}$}
\begin{eqnarray}
  d_{2\x34\x1}(1^+,2^+,3^+,4^-) &=& \mC_{2\x3\x4\x1}^{(3)}\,\ehat_{\{2^+\x3^+\x4^-\x1^+\}} \\ 
 &+& \frac{\spa2.4}{\spa1.2 \spa2.3} \Biggl[
   2 \frac{\spa1.4 \spa2.4 \spab1.(3+4).2 \spab2.(3+4).1}{\spa1.2^2 \spa3.4}
    +\frac{\spab4.(1+3).2 \spab4.(2+3).1}{2 \spa3.4}\nonumber \\ 
 &+&\frac{s_{1234} \spb1.3 \spb2.3}{2 \spb3.4}
    -2 m^2 \left( 3 \frac{\spa1.4 \spa2.4 \spb1.2}{\spa1.2 \spa3.4} 
                  +2 \frac{\spb1.3 \spb2.3}{\spb3.4}
                  +\frac{\spa2.4 \spb1.4 \spb2.3}{\spa2.3 \spb3.4} \right) \Biggr] \nonumber
\end{eqnarray}
\subsubsection{$d_{4\x3\x21}$}
\begin{eqnarray}
  d_{4\x3\x21}(1^+,2^+,3^+,4^-) &=& \mC_{1\x2\x3\x4}^{(2)} \, \ehat_{\{1^+\x2^+\x3^+\x4^-\}}  \nonumber \\
                               &+&(4\,m^2-s_{123})\left( \frac{s_{34}\,s_{123}^2}{2\,\spa1.2\,\spa2.3\,\spab1.(2+3).4\,\spab3.(1+2).4}\right)
\end{eqnarray}

\subsubsection{$d_{1\x23\x4}$}
\begin{eqnarray}
  d_{1\x23\x4}(1^+,2^+,3^+,4^-) =  \mC_{1\x2\x3\x4}^{(3)} \, \ehat_{\{1^+\x2^+\x3^+\x4^-\}} 
\end{eqnarray}

\subsubsection{$d_{2\x3\x4}$}
\begin{eqnarray}
  d_{2\x3\x4}(1^+,2^+,3^+,4^-)  &=& \mC_{1\x2\x3\x4}^{(1)}\,\ehat_{\{1^+\x2^+\x3^+\x4^-\}} + \mC_{2\x3\x4\x1}^{(5)}\,\ehat_{\{2^+\x3^+\x4^-\x1^+\}}\nonumber\\
                                &+& (4m^2s_{234}-s_{23}s_{1234}) \, \frac{\spa3.4 \spb2.3^2}{2\spa2.3\spab1.(3+4).2\spab1.(2+3).4}				
\end{eqnarray}

\subsubsection{$d_{1\x2\x3}$}
\begin{eqnarray}
  d_{1\x2\x3}(1^+,2^+,3^+,4^-) &=& \mC_{1\x2\x3\x4}^{(5)} \, \ehat_{\{1^+\x2^+\x3^+\x4^-\}} + \mC_{4\x1\x2\x3}^{(1)} \, \ehat_{\{4^-\x1^+\x2^+\x3^+\}} \nonumber \\
                               &+& (4\,m^2-s_{123})\, \frac{s_{123}\,\spb1.2\,\spb2.3}{2\,\spab3.(1+2).4\,\spab1.(2+3).4}
\end{eqnarray}
Note that, as expected, this whole expression is invariant under $1 \leftrightarrow 3$ since
$\mC_{4\times1\times2\times3}^{(1)} = \mC_{3\times2\times1\times4}^{(5)}$.

\subsubsection{$d_{3\x4\x1}$}
\begin{eqnarray}
  d_{3\x4\x1}(1^+,2^+,3^+,4^-) &=& \mC_{2\x3\x4\x1}^{(1)}\,\ehat_{\{2^+\x3^+\x4^-\x1^+\}} + \mC_{3\x4\x1\x2}^{(5)}\,\ehat_{\{3^+\x4^-\x1^+\x2^+\}} \nonumber \\
                               &-& \frac{1}{2}\frac{s_{14}\spb2.3\spa3.4}{\spa1.3\spa2.3^2\spab1.(3+4).2}(\spab4.(1+3).2\spa2.3+4m^2\spa3.4) \nonumber  \\
                               &-& \frac{1}{2}\frac{\spb1.2\spa1.4s_{34}}{\spa1.2^2\spa1.3\spab3.(1+4).2}(\spab4.(1+3).2\spa2.1+4m^2\spa1.4) \nonumber  \\
                               &+& \frac{1}{2}\frac{\spa1.4\spa3.4}{\spa1.2\spa1.3\spa2.3}\Bigg[ 4\frac{s_{14}s_{34}}{\spa1.3} + \spb1.3 \big(s_{123} -12m^2 \big) \Bigg]
\end{eqnarray}
Note that this is manifestly symmetric under the exchange $1 \leftrightarrow 3$.

\subsection{Triangles}
For the case of the triangle coefficients we divide the coefficient 
into two pieces, to separate the mass dependence, see Eq.~(\ref{c0andc2}). 
For many coefficients the $c_{i \x j}^{(2)}$ term is equal to zero. For these cases, the full result is given by the 
$m^0$ term and we shall omit the superscript.

\subsubsection{$c_{3\x4}$}
\begin{equation}
c_{3\x4}(1^+,2^+,3^+,4^-)= 2 s_{34}\,
\frac{\spa1.4\spa4.3}{\spa1.2 \spa2.3 \spa1.3^2},\;\;\;
\end{equation}

\subsubsection{$c_{2\x34}$}
\begin{eqnarray}
c_{2\x34}(1^+,2^+,3^+,4^-)&=&
 -2 (s_{23}+s_{24}) \frac{\spa1.4\,\spa2.4^2 }{\spa1.2^3\,\spa2.3\,\spa3.4}
\end{eqnarray}

\subsubsection{$c_{1\x43}$}
\begin{equation}
c_{1\x43}(1^+,2^+,3^+,4^-)= -2 (s_{13}+s_{14}) \; 
\frac{(\spa1.2\spa3.4+\spa1.3\spa2.4)}{\spa1.2\spa3.4}\Big[\frac{\spa1.4}{\spa1.2\spa1.3}\Big]^2, \;\;\;
\end{equation}

\subsubsection{$c_{4\x123}^{(0)}, c_{4\x123}^{(2)}$}
\begin{eqnarray}
c^{(0)}_{4\x123}(1^+,2^+,3^+,4^-)&=&-\frac{s_{123}^2\, (s_{14}+s_{24}+s_{34})}
   {\spa1.2\,\spa2.3\,\spab3.(1+2).4\,\spab1.(2+3).4}
\end{eqnarray}

\begin{eqnarray}
c^{(2)}_{4\x123}(1^+,2^+,3^+,4^-)&=&4 \frac {s_{123}\,(s_{14}+s_{24}+s_{34})}
   {\spa1.2\,\spa2.3\,\spab3.(1+2).4\,\spab1.(2+3).4}
\end{eqnarray}
\subsubsection{$c_{1\x234}^{(0)}, c_{1\x234}^{(2)}$}

\begin{eqnarray}
 &&c^{(0)}_{1\x234}(1^+,2^+,3^+,4^-)= 
  -2 \frac{\spab1.(2+3).1^2 \spab4.(2+3).1}{\spab1.(2+3).4 \spa1.2 \spa2.3 \spa3.4 \spb1.4}  \nonumber \\
  &&+(s_{12}+s_{13}+s_{14})
 \Bigg[ \frac{1}{\spab1.(2+3).4} \left(
   2 \frac{\spab4.(2+3).1 \spb2.3^2}{s_{34} s_{234}}
  -\frac{(s_{12}+s_{13})^2 \spb3.4+s_{23} \spa1.2 \spb2.3 \spb1.4}
   {\spa1.2 \spa1.3 \spa2.3 \spb1.4 \spb3.4} \right)
  \nonumber \\ &&
  +\frac{\spab3.(1+4).2 \spa1.4 \spb2.3^2}
  {\spab1.(3+4).2 \spa1.2 \spa1.3 \spb2.4 s_{34}}
  -\frac{\spab4.(2+3).1 \spb1.3^2}
  {\spab2.(3+4).1 \spa1.2 \spb1.4 s_{34}}
  -\frac{\spa1.4 (s_{12}^2 \spa2.3+2 \spa1.3 \spa2.4 \spb1.4 s_{24})}
  {\spa1.2^3 \spa2.3 \spa1.3 \spa3.4 \spb2.4 \spb1.4} \Bigg]
 \nonumber \\ 
\end{eqnarray}

\begin{eqnarray}
c_{1\x234}^{(2)}(1^+,2^+,3^+,4^-)&=& \frac{8\,\spab4.(2+3).1^3}
      {s_{234}\,(s_{12}+s_{13}+s_{14})\,\spa2.3\,\spa3.4\,\spab2.(3+4).1} \nonumber \\
     &-& 4\,(s_{12}+s_{13}+s_{14}) 
     \Bigg\{\frac{\spa1.4^2\,\spb2.3}{\spa1.2^2\,\spa3.4\,\spab1.(3+4).2\,\spab1.(2+3).4} \nonumber \\
       &-&\frac{\spb2.3}{\spa1.2\,\spab1.(2+3).4\,\spab1.(3+4).2\,\spab2.(3+4).1} 
       \Big[\frac{\spa1.2\spb1.3\spb2.3}{\spb3.4}
       +\frac{\spb1.2\spa1.4\spa2.4}{\spa3.4}\Big] \nonumber \\
     &-& \Big[\frac{\spb1.3\,\spa1.4}{\spa1.2}+\frac{\spab4.(2+3).1}{\spa2.3}\Big] 
      \Big[\frac{\spa2.4}{\spa1.2\,\spa3.4\,\spab1.(2+3).4\,\spab2.(3+4).1}\Big]\Bigg\}
\end{eqnarray}
\subsubsection{$c_{2\x341}^{(0)}, c_{2\x341}^{(2)}$}
\begin{eqnarray}
&&c_{2\x341}^{(0)}(1^+,2^+,3^+,4^-)= 
2\frac{(s_{12}+s_{23}+s_{24}) \spa2.4^2 \big(\spa1.4^2\spa2.3^2+\spa1.2^2\spa3.4^2\big)}
    {\spa1.2^3\spa2.3^3\spa1.4\spa3.4} \nonumber \\
     &+& \frac{\spab4.(1+3).2^4}{s_{134} \spa1.4 \spa3.4 \spab1.(3+4).2 \spab3.(1+4).2} 
     + \frac{(s_{12}+s_{23})\spb1.3^2}{\spa1.2 \spb1.4 \spa2.3 \spb3.4} \nonumber \\
     &+& \symbrack\left\{\frac{(s_{12}+s_{23}+s_{24}) \spb1.2 \spb1.3^3 }{s_{134} \spb1.4 \spb3.4 \spab2.(3+4).1}
     - \frac{\spb1.2 \spa2.4^2 \spab4.(3+2).1}{\spa1.2 \spa2.3 \spa3.4 \spab2.(3+4).1} 
     - \frac{\spb1.2 \spb1.3^2 \spa2.4}{\spa2.3 \spb3.4 \spab2.(3+4).1}\symbrack\right\} \nonumber \\
     &+& \symbrack\{1 \leftrightarrow 3 \symbrack\}
\end{eqnarray}

\begin{eqnarray}
c_{2\x341}^{(2)}(1^+,2^+,3^+,4^-)&=& 
        \frac{4 (s_{12}+s_{23}+s_{24})\,s_{134}\, \spb1.2\,\spb3.2}
      {\spa1.2\,\spa3.2\,\spb1.4\,\spb3.4\,\spab1.(3+4).2\,\spab3.(1+4).2}\nonumber \\
 &-& \frac{8}{(s_{12}+s_{23}+s_{24})\,s_{134}}
      \frac{\spab4.(1+3).2^4}{\spa1.4\,\spa3.4\,\spab1.(3+4).2\,\spab3.(1+4).2} \nonumber \\
  &+& \symbrack\Bigg\{\frac{4\,(s_{12}+s_{23}+s_{24})}{[(s_{13}+s_{14})\,(s_{23}+s_{24})-s_{12}\,s_{34}]}
      \Big[\frac{\spb1.3^2\,\spb2.3}{\spa1.2\,\spb1.4\,\spb3.4}
      -\frac{\spb1.2\,\spa1.4\,\spa2.4^2}{\spa1.2^2\,\spa2.3\,\spa3.4}\Big]\symbrack\Bigg\} \nonumber \\
  &+& \symbrack\Big\{ 1 \leftrightarrow 3\symbrack\Big\}
\end{eqnarray}
\subsubsection{$c_{12\x34}^{(0)}, c_{12\x34}^{(2)}$}
\begin{eqnarray}
     c^{(0)}_{12\x34}(1^+,2^+,3^+,4^-) &=&
        \frac{\spb1.3\,(s_{12}+s_{23})\,\spab4.(2+3).1}{\spa1.2\,\spa2.3\,\spb2.4\,\spab2.(3+4).1} 
      + \frac{\spb1.2\,\spa2.4^2\,\spab4.(2+3).1}{\spa1.2\,\spa2.3\,\spa3.4\,\spab2.(3+4).1} \nonumber \\
     &-& \frac{\spb2.3\,\spab4.(1+3).2^2}{\spa1.2\,\spb2.4\,\spa3.4\,\spab1.(3+4).2} 
      - \frac{\spb2.3^3\,s_{1234}}{\spa1.2\,\spb2.4\,\spb3.4\,\spab1.(3+4).2}   \nonumber \\
     &-& \frac{\spb1.3^2\,\spb2.3\,s_{1234}}{\spa1.2\,\spb2.4\,\spb3.4\,\spab2.(3+4).1} 
      + \frac{\spb1.2\,s_{123}\,(s_{123}-s_{124})}{\spa1.2\,\spa2.3\,\spb2.4\,\spab3.(1+2).4} \nonumber \\
     &-& \frac{\spb1.2\,\spa1.4\,\spab4.(2+3).1}{\spa1.2\,\spa2.3\,\spb2.4\,\spa3.4}
\end{eqnarray}

\begin{eqnarray}
c_{12\x34}^{(2)}(1^+,2^+,3^+,4^-) &=&  
  -4\,\frac{\spb1.2}{\spa1.2\,\spab3.(1+2).4\,\big[(s_{13}+s_{14})\,(s_{23}+s_{24})-s_{12}\,s_{34}\big]}
\nonumber \\
&\times&\Bigg[
   \left(\spa1.4\,\spb1.3-\spa2.4\,\spb2.3\right) (s_{13}+s_{23}-s_{14}-s_{24}) \nonumber \\ 
  &+&\spa1.2\,\spa3.4\,\spb1.3\,\spb2.3\,\left(2-\frac{(s_{13}+s_{23})\,(s_{13}+s_{14}+s_{23}+s_{24})}{s_{12}\,s_{34}}\right) \nonumber \\
  &+&\spa1.4\,\spa2.4\,\spb1.2\,\spb3.4\,\left(2-\frac{(s_{14}+s_{24})\,(s_{13}+s_{14}+s_{23}+s_{24})}{s_{12}\,s_{34}}\right)\Bigg]
\end{eqnarray}

\subsection{Bubbles}
\subsubsection{$b_{34}$}
\begin{equation}
b_{34}(1^+,2^+,3^+,4^-) = \frac{4}{\spa1.2^2 \spa1.3 \spa2.3} \left(
         \frac{\spa2.4^2\spa1.3\spb2.3}{(s_{23}+s_{24})}
       - \frac{\spa1.4^2\spa2.3\spb1.3}{(s_{13}+s_{14})} \right)
\end{equation}
\subsubsection{$b_{234}$}
\begin{eqnarray}
b_{234}(1^+,2^+,3^+,4^-) &=& \frac{4}{\spa2.3\, \spa3.4} \Bigg(
          \frac{\spa2.4^2\,\spab4.(2+3).1}{\spa1.2^2\,\spab2.(3+4).1} \nonumber \\
       &-&  \frac{\spab4.(2+3).1^3}{\spab2.(3+4).1\,(s_{1234}-s_{234})^2}
       - \frac{\spa2.4^2\,\spb2.3\,\spa3.4}{\spa1.2^2\,(s_{23}+s_{24})} \Bigg)
\end{eqnarray}
\subsubsection{$b_{1234}$}
Since the full amplitude is finite in four dimensions, one of the coefficients
is uniquely determined in terms of the remainder.  We thus have,
\begin{equation}
\label{b1234relation}
b_{1234}(1^+,2^+,3^+,4^-) = -b_{34}-b_{41}-b_{234}-b_{412}-b_{341} \,,
\end{equation}
where we have suppressed momentum and helicity labels on the right-hand side
for brevity.

\subsection{Rational terms} 
\begin{eqnarray}
r(1^+,2^+,3^+,4^-) &=& \frac{1}{2}\big[c_{12\x34}^{(2)}(1^+,2^+,3^+,4^-)+c_{12\x34}^{(2)}(3^+,2^+,1^+,4^-)\nonumber \\
&+&c_{1\x234}^{(2)}(1^+,2^+,3^+,4^-)+c_{1\x234}^{(2)}(3^+,2^+,1^+,4^-)\nonumber \\
&+&c_{4\x123}^{(2)}(1^+,2^+,3^+,4^-)+c_{2\x341}^{(2)}(1^+,2^+,3^+,4^-)\big]
\end{eqnarray}

\section{Coefficients for $H^{1234}_4(g^+,g^-,g^+,g^-;\Higgs)$}
For this helicity combination the coefficients of the scalar pentagon integrals contain a factor
of $1/\Trfourgfive1.2.3.4^4$ and we must modify the pentagon integral
coefficients in a similar fashion as described for the $+++-$ configuration
in Section~\ref{pppm}.  All coefficients 
can then be written in terms of a single function,
\begin{equation}
      \ehat_{\{1^+\x2^-\x3^+\x4^-\}}= -m^2 \frac{\spa1.2\spb3.4\spab4.(2+3).1}{\spb1.2\spa3.4\spab1.(2+3).4}
                                    \Bigg( \frac{\spb1.3^2\spa3.4}{\spb3.4} + \frac{\spa2.4^2\spb1.2}{\spa1.2} + 4 m^2 \frac{\spab4.(2+3).1}{\spab1.(2+3).4}\Bigg)
\end{equation}
This is manifestly symmetric under $\bigl\{ 1 \leftrightarrow 4, \, 2 \leftrightarrow 3, \langle \rangle \leftrightarrow [] \bigr\}$.
Other coefficients are trivially related via symmetries:
\begin{eqnarray}
\ehat_{\{3^+\x4^-\x1^+\x2^-\}}&=& \ehat_{\{1^+\x2^-\x3^+\x4^-\}}
 \Bigl\{ 1 \leftrightarrow 3, \, 2 \leftrightarrow 4 \Bigr\} \\
\ehat_{\{4^-\x1^+\x2^-\x3^+\}}&=& \ehat_{\{1^+\x2^-\x3^+\x4^-\}}
 \Bigl\{ 1 \to 4, \, 2 \to 1, \, 3 \to 2, \, 4 \to 3, \,
 \langle \rangle \leftrightarrow [] \Bigr\} \\
\ehat_{\{2^-\x3^+\x4^-\x1^+\}}&=& \ehat_{\{1^+\x2^-\x3^+\x4^-\}}
 \Bigl\{ 1 \to 2, \, 2 \to 3, \, 3 \to 4, \, 4 \to 1, \,
 \langle \rangle \leftrightarrow [] \Bigr\}
\end{eqnarray}
The minimal set of integral coefficients that must be calculated for the colour ordering $H_4^{1234}$
is shown in the first and third columns of Table~\ref{table:pmpm}, for example the bubble coefficients are given by:
\begin{eqnarray}
b_{412}(1^+,2^-,3^+,4^-) &=& b_{234}(3^+,4^-,1^+,2^-) \nonumber \\
b_{123}(1^+,2^-,3^+,4^-) &=& b_{234}(4^+,1^-,2^+,3^-)|_{\langle\,\rangle\leftrightarrow[\,]} \nonumber \\
b_{341}(1^+,2^-,3^+,4^-) &=& b_{234}(2^+,3^-,4^+,1^-)|_{\langle\,\rangle\leftrightarrow[\,]} \,.
\end{eqnarray}
The calculation of the coefficients of other colour orderings
requires the use of $+\,+\,-\,-$ functions which are given in the next section. 
\begin{table}
\begin{center}
\begin{tabular}{|l|l||l|l|}
\hline
Coefficient      & Related coefficients                                    &   Coefficient    & Related coefficients \\
\hline
  $d_{4\x3\x21}$ & $d_{2\x1\x43},d_{3\x2\x14},d_{1\x4\x32},$               &   $c_{3\x4} $    & $c_{4\x1},c_{2\x3},c_{1\x2}$\\               
                 & $d_{1\x2\x34},d_{2\x3\x41},$                            &   $c_{2\x34}$    & $c_{3\x41},c_{4\x12},c_{1\x23}$\\            
                 & $d_{3\x4\x12},d_{4\x1\x23}$                             &                  & $c_{1\x43},c_{2\x14},c_{3\x21},c_{4\x32}$\\ 
 $d_{1\x23\x4}$ & $d_{2\x34\x1},d_{3\x41\x2},d_{4\x12\x3}$                 &   $c_{12\x34}$   & $c_{23\x41}$ \\                              
  $d_{1\x2\x3}$  & $d_{2\x3\x4},d_{3\x4\x1},d_{4\x1\x2}$                   &   $c_{1\x234}$   & $c_{2\x341},c_{3\x412},c_{4\x123}$\\         
                 &                                                         &   $b_{34}$       & $b_{12},b_{23},b_{41}$ \\                    
                 &                                                         &   $b_{234}$      & $b_{341},b_{412},b_{123}$ \\                 
                 &                                                         &   $b_{1234}$     &  \\                 
\hline
\end{tabular}
\caption{Minimal set of integral coefficients for $1_g^+\,2_g^-\,3_g^+\,4_g^-$.}
\label{table:pmpm}
\end{center}
\end{table}

\subsection{Boxes}
\subsubsection{$d_{4\x3\x21}$}
\begin{eqnarray}
  d_{4\x3\x21}(1^+,2^-,3^+,4^-) &=& \ehat_{\{1^+\x2^-\x3^+\x4^-\}} \,\mC_{1\x2\x3\x4}^{(2)} \nonumber \\
  &+&\frac{\spab2.(1+3).4}{\spab1.(2+3).4\spab3.(1+2).4} \Bigg[
  -2\frac{\spa2.3\spab2.(1+3).4s_{34}s_{123}^2}{\spa1.2\spab3.(1+2).4^2} \nonumber \\
  &+&\frac{1}{2}\frac{\spa2.4^2\spb3.4s_{123}}{\spa1.2}+\frac{1}{2}\frac{\spb1.3^2\spa3.4s_{123}}{\spb1.2} -2m^2 \Bigg(2\frac{\spb1.3\spab4.(2+3).1}{\spb1.2} \nonumber \\
  &+&\frac{\spb2.3\spab2.(1+3).4\spab4.(2+3).1}{\spb1.2\spab1.(2+3).4}
  +3\frac{\spa2.3\spab2.(1+3).4\spab4.(1+2).3}{\spa1.2\spab3.(1+2).4} \Bigg) \Bigg]
\end{eqnarray}
\subsubsection{$d_{1\x23\x4}$}
\begin{eqnarray}
  d_{1\x23\x4}(1^+,2^-,3^+,4^-) &=& \ehat_{\{1^+\x2^-\x3^+\x4^-\}} \,\mC_{1\x2\x3\x4}^{(3)} \nonumber \\
                                &+& \symbrack\Bigg[\frac{1}{2} \frac{\spa2.4^3\spab4.(2+3).1}{\spa2.3\spa3.4\spa4.1} \Bigg( 1 + 4 m^2 \frac{\spa1.2}{\spa2.4\spab1.(2+3).4}\Bigg)\symbrack\Bigg] \nonumber \\
                                &+& \symbrack\Big[1 \leftrightarrow 4, \, 2 \leftrightarrow 3, \langle \rangle \leftrightarrow [] \symbrack\Big]
\end{eqnarray}
Note that here the symmetrization only applies to the terms in square brackets.
\subsubsection{$d_{1\x2\x3}$}
\begin{eqnarray}
  d_{1\x2\x3}(1^+,2^-,3^+,4^-) &=& \mC_{1\x2\x3\x4}^{(5)} \,
                                   \ehat_{\{1^+\x2^-\x3^+\x4^-\}} +
                                   \mC_{4\x1\x2\x3}^{(1)} \, \ehat_ {\{4^-\x1^+\x2^-\x3^+\}} \nonumber \\
  &+&\frac{\spa1.2\spa2.3}{\spab1.(2+3).4\spab3.(1+2).4} \Bigg[ 
      -2\frac{s_{12}s_{23}s_{123}}{\spa1.3^2}
      - \frac{1}{2} \frac{\spb1.2\spb2.3\spa2.4^2(s_{14}+s_{24}+s_{34})}{\spa1.4\spa3.4} \nonumber \\
  &+& \frac{1}{2} \spb1.3^2s_{123} + 2m^2 \Bigg(2\frac{\spb1.3s_{123}}{\spa1.3} -\spb1.3^2
      + \frac{\spb1.2\spb2.3\spa2.4^2}{\spa1.4\spa3.4}  \nonumber \\
  &-&\frac{\spb2.3\spab2.(1+3).4\spab4.(2+3).1}{\spa3.4\spab1.(2+3).4}
      + \frac{\spb1.2\spab2.(1+3).4\spab4.(1+2).3}{\spa1.4\spab3.(1+2).4} \Bigg) \Bigg]
\end{eqnarray}
Note that this is explicitly symmetric under the exchange $1 \leftrightarrow 3$.

\subsection{Triangles}
\subsubsection{$c_{3\x4}$}
\begin{eqnarray}
&&c_{3\x4}(1^+,2^-,3^+,4^-)=\frac{-2 s_{34}}{s_{12} \spab3.(1+2).4^3 \spa1.3^2 \spb2.4^2} \nonumber \\
&& \left\{ \spab3.1.4^3 \left[s_{24}(s_{14}+s_{24})+s_{12}(s_{23}+s_{34})\right] + 
 \spab3.1.4^2 \spab3.2.4 \left[s_{12}^2-s_{14}(s_{14}+s_{24})+s_{12}(3s_{234}-5s_{24}) \right] \right. \nonumber \\
&& \left. + \spab3.2.4^3 \left[s_{13}(s_{13}+s_{23})+s_{12}(s_{14}+s_{34})\right] + 
 \spab3.1.4 \spab3.2.4^2 \left[s_{12}^2-s_{23}(s_{13}+s_{23})+s_{12}(3s_{134}-5s_{13}) \right] \right\} \nonumber \\
\end{eqnarray}

\subsubsection{$c_{2\x34}$}
\begin{eqnarray}
c_{2\x34}(1^+,2^-,3^+,4^-) &=& 2 (s_{23}+s_{24}) 
\frac{s_{234}\,\spb2.3^2 \Big(\spab1.(2+4).3 \, \spb2.4+\spab1.(3+4).2 \, \spb3.4\Big)}
{\spab1.(3+4).2^3\,\spb2.4^2\,\spb3.4}
\end{eqnarray}
 
\subsubsection{$c_{12\x34}^{(0)},c_{12\x34}^{(2)}$}
This coefficient is defined in terms of the corresponding coefficient with a scalar loop, $\tilde{c}^{(0)}_{12\x34}$
\begin{eqnarray}
c^{(0)}_{12\x34}(1^+,2^-,3^+,4^-) &=& \tilde{c}^{(0)}_{12\x34}(1^+,2^-,3^+,4^-) \nonumber \\
     &+&\symbrack\Bigg\{\frac{\spa2.1^2\,\spb1.3^2\,\spa3.4^2-\spa2.4^2\,\spab1.(3+4).1 \,\spab4.(1+2).4}
     {\spa1.2\,\spa3.4\,\spab1.(3+4).2\,\spab3.(1+2).4} \symbrack\Bigg\} \\
   &+&\symbrack\Big\{ 1\leftrightarrow 3, 2\leftrightarrow 4 \symbrack\Big\}
    + \symbrack\Big\{ 1\leftrightarrow 2, 3\leftrightarrow 4, \langle \,\rangle \leftrightarrow [\,] \symbrack\Big\}  
    + \symbrack\Big\{ 1\leftrightarrow 4, 2\leftrightarrow 3, \langle \,\rangle \leftrightarrow [\,] \symbrack\Big\}   \nonumber 
\end{eqnarray}
\begin{eqnarray}
c_{12\x34}^{(2)}(1^+,2^-,3^+,4^-)
 &=& \symbrack\Bigg\{ 4 \frac{\spab2.(3+4).1}{\spab1.(3+4).2 \spab3.(1+2).4}  \Biggl[ 
    \frac{\spb2.3^2 \left(s_{23}-s_{14}\right)}{\spb1.2 \spb3.4 \spab1.(3+4).2}
   +\frac{3}{2} \frac{\spb1.3 \spb2.3}{\spb1.2 \spb3.4} \nonumber \\
 &+&\spa2.4 \frac{\left(\spab3.(1+2).3-\spab4.(1+2).4\right)}{\DeltaThree(1,2,3,4)} 
   \left(\spb2.3-\frac{\spa1.4 p_{12} \cdot p_{34}}{\spa1.2 \spa3.4}\right) \Bigg]\symbrack\Bigg\} \\
&+& \symbrack\Big\{1\leftrightarrow3,2\leftrightarrow4\symbrack\Big\}
+ \symbrack\Big\{1\leftrightarrow2,3\leftrightarrow4,\langle\,\rangle\leftrightarrow[\,]\symbrack\Big\}
+ \symbrack\Big\{1\leftrightarrow4,2\leftrightarrow3,\langle\,\rangle\leftrightarrow[\,]\symbrack\Big\} \nonumber
\end{eqnarray}
where $\DeltaThree$ is given by Eq.~(\ref{Delta3eqn}).

The coefficient for a scalar loop is,
\begin{eqnarray} \label{scpmpmC12x34}
&&\tilde{c}_{12\x34}^{(0)}(1^+,2^-,3^+,4^-)
     = \symbrack\Bigg\{ 
     2\,\frac{\spa2.3^3\,\spb3.4\,\spab3.(1+2).3\,(\spab3.(1+2).3\,\spb2.3-\spb1.2\,\spa1.4\,\spb3.4)}{\spa1.2\,\spab1.(3+4).2\,\spab3.(1+2).4^3} 
\nonumber \\
     &+&2\,\frac{\spa2.3^2\,\spb3.4\,\spab4.(1+2).3\,(-2\,s_{23}-s_{24})}{\spa1.2\,\spab1.(3+4).2\,\spab3.(1+2).4^2}
\nonumber \\
     &+&2\,\frac{\spb1.2\spa2.3^2\spb3.4
      (2s_{12}(s_{23}-s_{14}-s_{34})+2s_{13}s_{23}
      +2s_{23}^2+s_{14}s_{34}-s_{23}s_{34}
      +2\spb1.2\spa1.3\spa2.4\spb3.4)}{\spab1.(3+4).2\spab3.(1+2).4^3}
\nonumber \\
     &+& 2\,\frac{\spb1.4^2(\spa1.4\spa2.4(2(s_{13}-s_{24})-3(s_{34}+s_{14})-4(s_{12}+s_{23}))
       -2\spa1.3\spb2.3\spa2.4^2
       +3\spb1.3\spa1.4^2\spa2.3)}{\spab1.(3+4).2\spab3.(1+2).4^2}
\nonumber \\
     &+& \frac{s_{14}^2\,s_{12}\,(6\,s_{13}-2\,s_{14}+2\,s_{23}+2\,s_{24})-s_{14}^4+s_{14}^2\,s_{23}^2}{\spab1.(3+4).2^2\,\spab3.(1+2).4^2}
     - 8\,\frac{s_{12}\,s_{13}\,s_{14}\,s_{23}}{\spab1.(3+4).2^2\,\spab3.(1+2).4^2}
\nonumber \\
     &+& 4\,\frac{s_{14}\,s_{1234}\,\spab2.(3+4).1\,\spab4.(1+2).3}{\spab1.(3+4).2\,\spab3.(1+2).4\,\DeltaThree(1,2,3,4)}
     + 4\,\frac{\spa1.2\,\spb1.3\,\spab2.(3+4).1\,\spab4.(1+2).3\,\spab3.(1+4).2}{\spab1.(3+4).2\,\spab3.(1+2).4\,\DeltaThree(1,2,3,4)}
\nonumber \\
     &+& \frac{\spab1.(2+3).4\,\spab2.(3+4).1\,\spab3.(1+4).2\,\spab4.(1+2).3\,(\Pi(4,3,2,1)\,\Pi(1,2,3,4)+\DeltaThree(1,2,3,4))}{\spab1.(3+4).2^2\,\spab3.(1+2).4^2\,\DeltaThree(1,2,3,4)}
\nonumber \\
     &-& 3\,\frac{s_{1234}\,\spab2.(3+4).1\,\spab4.(1+2).3\,\Pi(4,3,2,1)\,\Pi(1,2,3,4)\,(s_{13}+s_{14}+s_{23}+s_{24})}{2\,\spab1.(3+4).2\,\spab3.(1+2).4\,\DeltaThree(1,2,3,4)^2}
\nonumber \\
     &+& 5\,\frac{s_{1234}\,\spab2.(3+4).1\,\spab4.(1+2).3\,(s_{13}+s_{14}+s_{23}+s_{24})}{2\,\spab1.(3+4).2\,\spab3.(1+2).4\,\DeltaThree(1,2,3,4)}
       - 4\,\frac{\spab2.(3+4).1\,\spab4.(1+2).3}{\spab1.(3+4).2\,\spab3.(1+2).4} \symbrack\Bigg\}
\nonumber \\
&+& \symbrack\Big\{1\leftrightarrow3,2\leftrightarrow4\symbrack\Big\}
+ \symbrack\Big\{1\leftrightarrow2,3\leftrightarrow4,\langle\,\rangle\leftrightarrow[\,]\symbrack\Big\}
+ \symbrack\Big\{1\leftrightarrow4,2\leftrightarrow3,\langle\,\rangle\leftrightarrow[\,]\symbrack\Big\}
\end{eqnarray}
where
\begin{equation}
\Pi(i,j,k,l)=s_{ik}+s_{jk}-s_{il}-s_{jl}
\end{equation}
and $\DeltaThree$ is given by Eq.~(\ref{Delta3eqn}).

\subsubsection{$c_{1\x234}^{(0)},c_{1\x234}^{(2)}$}
\begin{eqnarray}
&&c_{1\x234}^{(0)}(1^+,2^-,3^+,4^-)=- 2\,\frac{\spb1.3^4}{\spb1.2\,\spb1.4\,\spb3.2\,\spb3.4} \nonumber \\
     &+&(s_{12}+s_{13}+s_{14})\,\Bigg[
     \frac{\spb1.3^2\,\spb3.4}{\spab1.(2+3).4\,\spb1.4\,\spb2.3\,\spb2.4} 
     +\frac{\spb1.3^2\,\spb3.2}{\spab1.(3+4).2\,\spb1.2\,\spb4.3\,\spb4.2} \nonumber \\
     &+&\frac{\spa2.4^2\,\spa1.4}{\spab1.(3+4).2\,\spa1.3\,\spa1.2\,\spa3.4} 
     +\frac{\spa2.4^2\,\spa1.2}{\spab1.(2+3).4\,\spa1.3\,\spa1.4\,\spa3.2} \nonumber \\
     &-&2\,\frac{s_{234}\,\spab1.(2+4).3^2\,\Big(\spb2.3^2\,\spab1.(2+3).4^2+\spb3.4^2\,\spab1.(3+4).2^2\Big)}
     {\spab1.(2+3).4^3\,\spab1.(3+4).2^3\,\spb2.3\,\spb3.4} \Bigg] \nonumber \\
\end{eqnarray}
\begin{eqnarray}
&&c_{1\x234}^{(2)}(1^+,2^-,3^+,4^-)=4\,\frac{(s_{12}+s_{13}+s_{14})}{\spab1.(3+4).2\,\spab1.(2+3).4}
     \,\Big[\Big(\frac{\spa2.4^2}{\spa2.3\,\spa3.4}-\frac{\spb1.3^2}{\spb1.2\,\spb1.4}\Big) \nonumber \\
     &+&\frac{\spab1.(2+4).3}{\spab1.(3+4).2}
     \,\Big(\frac{\spa1.4\,\spa2.4}{\spa1.2\,\spa3.4}+\frac{\spb1.3\,\spb2.3}{\spb1.2\,\spb3.4}\Big) 
     +\frac{\spab1.(2+4).3}{\spab1.(2+3).4}
       \,\Big(\frac{\spa1.2\,\spa2.4}{\spa1.4\,\spa2.3}+\frac{\spb1.3\,\spb3.4}{\spb1.4\,\spb2.3}\Big)\Big] \nonumber \\
     &-&8\,\frac{\spb1.3^4}{\spb1.2\,\spb1.4\,\spb2.3\,\spb3.4\,(s_{12}+s_{13}+s_{14})}    
\end{eqnarray}

\subsection{Bubbles}
\subsubsection{$b_{34}$}
\begin{eqnarray}
&&b_{34}(1^+,2^-,3^+,4^-) = 
 \symbrack\Bigg\{ 
  4\,\frac{\spb1.3\,\spa1.4^2\,s_{134}}{\spa1.3\,(s_{13}+s_{14})\,\spab1.(3+4).2^2} \nonumber \\
 &+&4\,\frac{\spb1.4^2\,\spa3.4\,(s_{13}+s_{14})\,\spab4.(1+2).3\,(2\,s_{123}+s_{124})}
    {\spb1.2\,\spab1.(3+4).2\,\spab3.(1+2).4^2\,\DeltaThree(1,2,3,4)} \nonumber \\ 
 &+&12\,\frac{\spa1.2\,\spb1.4^2\,\spa3.4\,\spab4.(1+2).3\,s_{134}}
    {\spab1.(3+4).2\,\spab3.(1+2).4^2\,\DeltaThree(1,2,3,4)}
 +4\,\frac{\spab4.(1+2).3\,\spab4.(1+3).4}{\spab1.(3+4).2^2\,\spab3.(1+2).4} \nonumber \\
 &+&8\,\frac{\spb1.3\,\spa3.4\,\spab1.(2+3).4\,\spab4.(1+2).3\,(s_{234}-s_{134})}
    {\spab1.(3+4).2^2\,\spab3.(1+2).4\,\DeltaThree(1,2,3,4)} \nonumber \\ 
 &-&2\,\frac{\spb1.4\,\spa2.3\,\spab4.(1+2).3}{\spab1.(3+4).2\,\spab3.(1+2).4^2}
 +4\,\frac{\spb1.4\,\spa2.3\,(s_{13}+s_{14})\,\spab4.(1+2).3\,(s_{123}-s_{124})}
    {\spab1.(3+4).2\,\spab3.(1+2).4^2\,\DeltaThree(1,2,3,4)} \nonumber \\
 &+&3\,\frac{\spab2.(3+4).1\,\spab4.(1+2).3\,(s_{123}-s_{124})\,(s_{234}-s_{134})\,(s_{134}+s_{234})}
    {\spab1.(3+4).2\,\spab3.(1+2).4\,\DeltaThree(1,2,3,4)^2} \nonumber \\
 &+&\frac{\spab2.(3+4).1\,\spab4.(1+2).3\,(s_{13}-5\,s_{14}-5\,s_{23}+s_{24}-14\,s_{34})}
    {\spab1.(3+4).2\,\spab3.(1+2).4\,\DeltaThree(1,2,3,4)}\symbrack\Bigg\}
 \nonumber \\
 &+& \symbrack\Big\{1 \leftrightarrow 2, 3 \leftrightarrow 4, \langle\,\rangle \leftrightarrow [\,] \symbrack\Big\} 
\end{eqnarray}
where $\DeltaThree$ is given by Eq.~(\ref{Delta3eqn}).
\subsubsection{$b_{234}$}
\begin{eqnarray}
b_{234}(1^+,2^-,3^+,4^-) &=& \symbrack\Bigg\{ \frac{4\,s_{234}\,\spb3.4}{\spb2.4} \bigg(
 -\frac{\spa2.4\,\spb3.4}{(s_{24}+s_{34})\,\spab1.(2+3).4^2} 
  \nonumber \\  
 &+&\frac{\spb1.3\,\spab1.(2+4).3}{(s_{1234}-s_{234})\,\spab1.(2+3).4\,\spb2.3\,\spb3.4}
     \left[ \frac{\spb1.3}{(s_{1234}-s_{234})}
      -\frac{\spb3.4}{\spab1.(2+3).4} \right]
  \bigg) \symbrack\Bigg\}\nonumber \\ 
  &+& \symbrack\Big\{2 \leftrightarrow 4\symbrack\Big\} 
\end{eqnarray}
\subsubsection{$b_{1234}$}
Similarly to Eq.~(\ref{b1234relation}) we have,
\begin{equation}
b_{1234}(1^+,2^-,3^+,4^-) =
 -b_{34}-b_{12}-b_{23}-b_{41} 
 -b_{234}-b_{341}-b_{412}-b_{123} \,,
\end{equation}
suppressing momentum and helicity labels on the right-hand side
for brevity.

\subsection{Rational terms} 
\begin{eqnarray}
r(1^+,2^-,3^+,4^-)&=&\frac{1}{2}\Big[c_{12\x34}^{(2)}(1^+,2^-,3^+,4^-) 
+c_{12\x34}^{(2)}(2^+,3^-,4^+,1^-)|_{[\,]\leftrightarrow \langle\,\rangle}  \nonumber \\
&+&c_{1\x234}^{(2)}(1^+,2^-,3^+,4^-)
+c_{1\x234}^{(2)}(2^+,3^-,4^+,1^-)|_{[\,]\leftrightarrow \langle\,\rangle} \nonumber \\
&+&c_{1\x234}^{(2)}(3^+,4^-,1^+,2^-) +c_{1\x234}^{(2)}(4^+,1^-,2^+,3^-)|_{[\,]\leftrightarrow \langle\,\rangle}\Big]
\end{eqnarray}

\section{Coefficients for $H^{1234}_4(g^+,g^+,g^-,g^-;\Higgs)$}
In this case, as for the all-plus helicity amplitude, there are no factors of $1/\Trfourgfive1.2.3.4^2$
in the pentagon integral coefficients.  Therefore the effective pentagon integral coefficients simply correspond
to the $\mu^2 \to 0$ limit, as in the $++++$ case.  We thus have,
\begin{eqnarray}
\ehat_{\{1^+\x2^+\x3^-\x4^-\}}
  &=& m^2 (s_{12}+s_{34}-4 m^2
)\frac{\spb1.2 \spa3.4}{\spa1.2 \spb3.4 } \\
\ehat_{\{2^+\x3^-\x4^-\x1^+\}} &=& m^2 \frac{\spa3.4}{\spb3.4} \Bigg[(s_{34}-4\,m^2)\frac{\spb2.3\,\spa3.4\,\spb4.1}{\spa2.3\,\spb3.4\,\spa4.1} 
-\spb1.2^2\Bigg] \\
  \ehat_{\{4^-\x1^+\x2^+\x3^-\}} &=& \ehat_{\{2^+\x3^-\x4^-\x1^+\}}\{2\leftrightarrow4, 1\leftrightarrow3, \langle\,\rangle\leftrightarrow[\,]\}
\end{eqnarray}
The  minimal set of integral coefficients that must be computed in this case
is shown in the first and third columns of Table~\ref{table:ppmm}.
For the colour ordering $H_4^{1234}$ the complete set of related coefficients
is given in Table~\ref{table:ppmm}, for example bubble coefficients are given by:
\begin{eqnarray}
b_{341}(1^+,2^+,3^-,4^-) &=& b_{234}(2^+,1^+,4^-,3^-) \nonumber \\
b_{123}(1^+,2^+,3^-,4^-) &=& b_{234}(4^+,3^+,2^-,1^-)|_{\langle\,\rangle\leftrightarrow[\,]}  \nonumber \\
b_{412}(1^+,2^+,3^-,4^-) &=& b_{234}(3^+,4^+,1^-,2^-)|_{\langle\,\rangle\leftrightarrow[\,]} \,.
\end{eqnarray}
The calculation of the coefficients of other colour orderings
requires the use of $+\,-\,+\,-$ functions which are given in the previous section. 
\begin{table}
\begin{center}
\begin{tabular}{|l|l||l|l|}
\hline
Coefficient          & Related coefficients                     &Coefficient          & Related coefficients \\
\hline
  $d_{1\x2\x34}$ & $d_{2\x1\x43},d_{3\x4\x12},d_{4\x3\x21} $    &  $c_{2\x3}$     & $c_{4\x1}$\\                           
  $d_{1\x4\x32}$ & $d_{3\x2\x14},d_{4\x1\x23},d_{2\x3\x41} $    &  $c_{1\x23}$    & $c_{2\x14},c_{3\x41},c_{4\x32}$  \\    
  $d_{2\x34\x1}$ & $d_{4\x12\x3}$                               &  $c_{23\x41}$   &  \\                        
  $d_{1\x23\x4}$ & $d_{3\x41\x2}$                               &  $c_{1\x234}$   & $c_{2\x341},c_{3\x412} ,c_{4\x123}$ \\ 
  $d_{1\x2\x3}$  & $d_{3\x4\x1},d_{4\x1\x2},d_{2\x3\x4}$        &  $b_{23}$       & $b_{41}$ \\                            
                 &                                              &  $b_{234}$      & $b_{341},b_{412},b_{123}$ \\           
                 &                                              &  $b_{1234}$     & \\           
\hline
\end{tabular}
\caption{Minimal set of integral coefficients for $1_g^+\,2_g^+\,3_g^-\,4_g^-$.}
\label{table:ppmm}
\end{center}
\end{table}

\subsection{Boxes}
\subsubsection{$d_{1\x2\x34}$}
\begin{eqnarray}
d_{1\x2\x34}(1^+,2^+,3^-,4^-)=\mC_{1\x2\x3\x4}^{(4)} \, \ehat_{\{1^+\x2^+\x3^-\x4^-\}}
\end{eqnarray}
\subsubsection{$d_{1\x4\x32}$}
\begin{eqnarray}
  d_{1\x4\x32}(1^+,2^+,3^-,4^-)&=& \mC_{2\x3\x4\x1}^{(2)} \,\ehat_ {\{2^+\x3^-\x4^-\x1^+\}} 
  -\frac{\spb2.4}{\spab1.(2+3).4\,\spb3.4}\nonumber \\
  &\times&\Bigg\{ 2\,\frac{\spb2.4\,s_{14}\,s_{234}^2 \spab1.(3+4).2}{\spb2.3\,\spab1.(2+3).4^2} 
  +\frac{1}{2}\,s_{234}\Big[\frac{\spb1.4\,\spa3.4^2}{\spa2.3}
                           +\frac{\spa1.4\,\spb1.2^2}{\spb2.3}\Big] \\
  &+&2\,m^2 \Big[3\,\frac{\spb2.4\spab1.(3+4).2\spab4.(2+3).1}{\spb2.3\spab1.(2+3).4}
  +\frac{\spb1.4\spa3.4s_{234}}{\spa2.3\spb3.4} 
  +\frac{\spa3.4\,\spab3.(2+4).1}{\spa2.3}\Big]\Bigg\}  \nonumber
\end{eqnarray}
\subsubsection{$d_{2\x34\x1}$}
\begin{eqnarray}
  d_{2\x34\x1}(1^+,2^+,3^-,4^-)&=& \mC_{2\x3\x4\x1}^{(3)} \,\ehat_{\{2^+\x3^-\x4^-\x1^+\}}
  +\frac{1}{2}\frac{\spa3.4\,\spab1.(3+4).2\,\spab2.(3+4).1\,(s_{34}-4m^2)}{\spa1.2\,\spa1.4\,\spa2.3\,\spb3.4^2}
  \nonumber \\
\end{eqnarray}
\subsubsection{$d_{1\x23\x4}$}
\begin{eqnarray}
d_{1\x23\x4}(1^+,2^+,3^-,4^-)&=&\mC_{1\x2\x3\x4}^{(3)}   \, \ehat_{\{1^+\x2^+\x3^-\x4^-\}}  \nonumber \\
&-& \frac{1}{2}\,\spab4.(2+3).1 \times 
\Bigg[ \frac{\spb2.1\,\spb2.4}{\spb1.4\,\spb2.3\,\spb3.4}\bigg(\spb2.1 - \frac{4\,m^2\,\spb2.4}{\spab1.(2+3).4}\bigg)\nonumber \\
&+&\frac{\spa4.3\,\spa1.3}{\spa2.3\,\spa1.4\,\spa1.2}\bigg(\spa4.3 - \frac{4\,m^2\,\spa1.3}{\spab1.(2+3).4}\bigg) \Bigg]
\end{eqnarray}
\subsubsection{$d_{1\x2\x3}$}
\begin{eqnarray}
d_{1\x2\x3}(1^+,2^+,3^-,4^-)&=& \mC_{1\x2\x3\x4}^{(5)} \,\ehat_{\{1^+\x2^+\x3^-\x4^-\}}
               +\mC_{4\x1\x2\x3}^{(1)}\,\ehat_{\{4^-\x1^+\x2^+\x3^-\}} \nonumber \\
&+& \frac{\spb1.2^2\,\spa2.3\,(s_{12} - 4\,m^2)}{2\,\spb3.4\,\spa1.2\,\spb1.4}
\end{eqnarray}

\subsection{Triangles}
\subsubsection{$c_{2\x3}$}
\begin{equation}
c_{2\x3}(1^+,2^+,3^-,4^-)=
 -2\, \frac{s_{23}}{\spab2.(1+4).3^3} 
   \Bigg\{\frac{\spb1.3^2\,s_{134} \, \spab2.(3+4).1}{\spb1.4\,\spb3.4}
               +\frac{\spa2.4^2\,s_{124} \,\spab4.(1+2).3}{\spa1.4\,\spa1.2}\Bigg\}
\end{equation}
\subsubsection{$c_{1\x23}$}
\begin{eqnarray}
  c_{1\x23}(1^+,2^+,3^-,4^-) &=& 2\,(s_{12}+s_{13})
                                       \frac{s_{123}\,\spa1.3^2\,\spab3.(1+2).4}{\spa1.2\,\spa2.3\,\spab1.(2+3).4^3} 
\end{eqnarray}
\subsubsection{$c_{23\x41}^{(0)}, c_{23\x41}^{(2)}$}
The full coefficient for this scalar integral is defined in terms of the 
coefficient with a scalar running in the loop, $\tilde{c}^{(0)}_{23\x41}$,
\begin{eqnarray}
c^{(0)}_{23\x41}(1^+,2^+,3^-,4^-)&=& \tilde{c}^{(0)}_{23\x41}(1^+,2^+,3^-,4^-) \nonumber \\
     &+&\symbrack\Bigg\{ -\frac{\spa3.2^2\,\spb2.1^2\,\spa1.4^2-\spa3.4^2\,\spab2.(1+4).2 \,\spab4.(2+3).4}
     {\spa2.3\,\spa1.4\,\spab2.(1+4).3\,\spab1.(2+3).4} \symbrack\Bigg\} \\
   &+&\symbrack\Big\{ 1\leftrightarrow 2, 3\leftrightarrow 4 \symbrack\Big\} 
   +\symbrack\Big\{ 1\leftrightarrow 3, 2\leftrightarrow 4, \langle \, \rangle \leftrightarrow [\,] \symbrack\Big\}
   +\symbrack\Big\{ 1\leftrightarrow 4, 2\leftrightarrow 3, \langle \, \rangle \leftrightarrow [\,] \symbrack\Big\} \nonumber
\end{eqnarray}
In turn the coefficient for a scalar loop is given by,
\begin{eqnarray}
  \tilde{c}_{23\x41}^{(0)}(1^+,2^+,3^-,4^-)&=& -\tilde{c}_{12\x34}^{(0)}(2^+,3^-,1^+,4^-) \\
&+& \symbrack\Bigg\{ -2\,\Delta(1,4,2,3) \Big[\frac{(s_{13}-s_{24})}{\spab2.(1+4).3\,\spab1.(2+3).4}\Big]^2
    -4\frac{\spab3.(1+4).2\,\spab4.(2+3).1}{\spab2.(1+4).3\,\spab1.(2+3).4}\symbrack\Bigg\} \nonumber
\end{eqnarray}
where $\tilde{c}_{12\x34}^{(0)}$ is given by Eq.~(\ref{scpmpmC12x34}) with the appropriate
permutation of arguments.
\begin{eqnarray}
c_{23\x41}^{(2)}(1^+,2^+,3^-,4^-) &=&
  \Biggl\{ 4 \frac{\spa2.4\,\spab3.(1+4).2}{\spa2.3\,\spa1.4\,\spab2.(1+4).3} \Bigg[
   \frac{1}{\spab1.(2+3).4}\left(\frac{\spa2.4\,(s_{13}-s_{24})}{\spab2.(1+4).3}-2\,\spa3.4 \right)
   \nonumber \\ &&
  -\frac{\spab4.(2+3).1}{\DeltaThree(1,4,2,3)} \left( \frac{\spa1.2\,\spab3.(1+4).2}{\spab1.(2+3).4}-\spa3.4 \right)
   \Bigg]\Biggr\} \\ 
&+& \symbrack\{ 1\leftrightarrow2,3\leftrightarrow4 \symbrack\}
  + \symbrack\{ 1\leftrightarrow3,2\leftrightarrow4,\langle\,\rangle\leftrightarrow[\,] \symbrack\}
  + \symbrack\{ 1\leftrightarrow4,2\leftrightarrow3,\langle\,\rangle\leftrightarrow[\,] \symbrack\} \nonumber
\end{eqnarray}
where $\DeltaThree$ is given by Eq.~(\ref{Delta3eqn}).

\subsubsection{$c_{1\x234}^{(0)}, c_{1\x234}^{(2)}$}
\begin{eqnarray}
c^{(0)}_{1\x234}(1^+,2^+,3^-,4^-)&=& 
  -2 (s_{12}+s_{13}+s_{14})\, s_{234}
     \frac{\spab1.(3+4).2\,\spb2.4^2}{\spab1.(2+3).4^3\,\spb2.3\,\spb3.4}
 +\frac{2\,\spb1.2^3}{\spb1.4\,\spb2.3\,\spb3.4} \nonumber \\
  &-&\frac{(s_{12}+s_{13}+s_{14})}{\spa1.2\,\spab1.(2+3).4\,\spb3.4}
     \left(\frac{\spa1.3\,\spa3.4^2\,\spb3.4}{\spa1.4\,\spa2.3}
          -\frac{\spa1.2\,\spb1.2^2\,\spb2.4}{\spb1.4\,\spb2.3}\right)
\end{eqnarray}
\begin{eqnarray}
c_{1\x234}^{(2)}(1^+,2^+,3^-,4^-) &=& \frac{-4}{s_{23} \spab1.4.3}
\Bigg\{\frac{\spab4.1.2^2(\spab3.4.1-\spab3.2.1)}{s_{14}(s_{12}+s_{13}+s_{14})} \nonumber \\
&& + \frac{\spb1.2\spa3.4}{\spab1.(2+3).4}\left(\spab1.(3+4).2+\frac{(s_{12}+s_{13}+s_{14})}{s_{12}}\spab1.3.2\right) \\
&& - 2\frac{\spab1.(3+4).2\spab4.(2+3).1\spab3.1.2}{(s_{12}+s_{13}+s_{14})\spab1.(2+3).4} 
+ \frac{\spab1.(3+4).2^2 \spab4.(2+3).1 \spab3.1.4}{(s_{12}+s_{13}+s_{14})\spab1.(2+3).4^2}\Bigg\} \nonumber
\end{eqnarray}

\subsection{Bubbles}
\subsubsection{$b_{23}$}
\begin{eqnarray}
&&b_{23}(1^+,2^+,3^-,4^-) = 
 \symbrack\Bigg\{ -4\,\frac{\spb2.4^2\,\spa3.4\,s_{234}}{\spb3.4\,\spab4.(2+3).4\,\spab1.(2+3).4^2} \nonumber \\
 &-&4\,\frac{\spb1.3^2\,\spa2.3\,\spab1.(2+3).1\,\spab3.(4+1).2\,(2\,s_{124}+s_{134}-3\,s_{14})}
   {\spb1.4\,\spab1.(2+3).4\,\spab2.(1+4).3^2\,\DeltaThree(1,4,2,3)} \nonumber \\ 
 &+&12 \frac{\spa1.4\,\spb1.3^2\,\spa2.3^2\,\spb2.3\,\spab3.(4+1).2}
   {\spab1.(2+3).4\,\spab2.(1+4).3^2\,\DeltaThree(1,4,2,3)}
 +4 \frac{\spa2.4\,\spb2.4\,\spab3.(4+1).2}{\spab1.(2+3).4^2\,\spab2.(1+4).3} \nonumber \\ 
 &-&8 \frac{\spb2.4\,\spa2.3\,\spab3.(4+1).2 \left(
      \spa1.4\,\spb3.4\,\spab4.(2+3).1
     +\spa1.4\,\spa2.3\,\spb1.3\,\spb2.3
     +\spa2.4\,\spb2.3\,\spab4.(2+3).4 \right)}
   {\spab1.(2+3).4^2\,\spab2.(1+4).3\,\DeltaThree(1,4,2,3)} \nonumber \\ 
 &-&4 \frac{\spa2.4\,\spb1.3\,\spab3.(4+1).2}{\spab1.(2+3).4\,\spab2.(1+4).3^2}
 +8 \frac{\spa1.4\,\spb3.4\,\spb1.3\,\spa2.3\,\spab4.(2+3).1\,\spab3.(4+1).2}
   {\spab1.(2+3).4\,\spab2.(1+4).3^2\,\DeltaThree(1,4,2,3)} \nonumber \\ 
 &+&3 \frac{\spab4.(2+3).1\,\spab3.(4+1).2\,(s_{124}-s_{134})\,(s_{123}-s_{234})\,(s_{234}+s_{123})}
   {\spab1.(2+3).4\,\spab2.(1+4).3\,\DeltaThree(1,4,2,3)^2} \nonumber \\ 
 &+&\frac{\spab4.(2+3).1\,\spab3.(4+1).2
    \left(2 s_{23}+5 s_{24}+3 s_{34}+3 s_{12}+5 s_{13}\right)}
   {\spab1.(2+3).4\,\spab2.(1+4).3\,\DeltaThree(1,4,2,3)} \symbrack\Bigg\} 
 \nonumber \\
  &+& \symbrack\Big\{1 \leftrightarrow 4, 2 \leftrightarrow 3, \langle \, \rangle \leftrightarrow [\,] \symbrack\Big\} 
 \end{eqnarray}
where $\DeltaThree$ is given by Eq~(\ref{Delta3eqn}).
\subsubsection{$b_{234}$}
\begin{eqnarray}
b_{234}(1^+,2^+,3^-,4^-) &=& \frac{4\,s_{234}}{\spb3.4\,\spab1.(2+3).4^2} \Bigg(
       \frac{\spa3.4\,\spb2.4^2}{(s_{24}+s_{34})} \nonumber \\ 
         &+&\frac{\spb1.2\,\spab1.(3+4).2}{\spb2.3} \Big[
	   \frac{\spab1.(2+3).4\,\spb1.2}{(s_{12}+s_{13}+s_{14})^2}
	  -\frac{\spb2.4}{(s_{12}+s_{13}+s_{14})} \Big] \Bigg)
\end{eqnarray}
\subsubsection{$b_{1234}$}
Similarly to Eq.~(\ref{b1234relation}) we have,
\begin{equation}
b_{1234}(1^+,2^+,3^-,4^-) = -b_{23}-b_{41}-b_{234}-b_{341}-b_{412}-b_{123} \,,
\end{equation}
suppressing momentum and helicity labels on the right-hand side
for brevity.

\subsection{Rational terms} 
\begin{eqnarray}
r(1^+,2^+,3^-,4^-)&=&\frac{1}{2}\Big[c_{23\x41}^{(2)}(1^+,2^+,3^-,4^-) \nonumber \\
 &+&c_{1\x234}^{(2)}(1^+,2^+,3^-,4^-)+c_{1\x234}^{(2)}(2^+,1^+,4^-,3^-) \nonumber \\
 &+&c_{1\x234}^{(2)}(3^+,4^+,1^-,2^-)|_{[\,]\leftrightarrow \langle\,\rangle} +c_{1\x234}^{(2)}(4^+,3^+,2^-,1^-)|_{[\,]\leftrightarrow \langle\,\rangle}  
\Big]
\end{eqnarray}

\section{Coefficients for $H^{34}_4(\qb^+,q^-,g^+,g^+;\Higgs)$}

The coefficients that must be computed for this amplitude are shown in
Table~\ref{table:aqgg}.
\renewcommand{\baselinestretch}{1.2}
\begin{table}
\begin{center}
\begin{tabular}{|l|l||l|l||l|l||}
\hline
\multicolumn{2}{|c||}{$1_{\qb}^+,2_q^-,3_g^+,4_g^+$}     & \multicolumn{2}{|c||}{$1_{\qb}^+,2_q^-,3_g^-,4_g^+$}  & \multicolumn{2}{|c||}{$1_{\qb}^+,2_q^-,3_g^+,4_g^-$}   \\
\hline
Coefficient     & Related             & Coefficient   & Related            & Coefficient  & Related \\
                & coefficients        &               & coefficients       &              & coefficients\\
\hline
$d_{3\x21\x4}$  &                     &$d_{3\x21\x4}$ &                    & $c_{4\x123}$ & $c_{3\x412}$ \\
$d_{4\x3\x21}$  & $d_{3\x4\x12}$      &$d_{4\x3\x21}$ & $d_{3\x4\x12}$     & $b_{123}$    & $b_{412}$ \\
$c_{3\x21}$     & $c_{4\x12}$         &$c_{3\x21}$    & $c_{4\x12}$        &              &\\
$c_{12\x34}$    &                     &$c_{3\x4}$     &                    &              &\\
$c_{4\x123}$    &                     &$c_{12\x34}$   &                    &              &\\
$c_{3\x412}$    &                     &$c_{4\x123}$   & $c_{3\x412}$       &              &\\
$b_{12}$        &                     &$b_{34}$       &                    &              &\\
$b_{123}$       &                     &$b_{12}$       &                    &              &\\
$b_{412}$       &                     &$b_{123}$      & $b_{412}$          &              &\\
$b_{1234}$      &                     &$b_{1234}$     &                    &              &\\
\hline
\end{tabular}
\caption{Minimal set of integral coefficients for $H^{34}_4(1_{\qb}^+,2_q^-,3_g^+,4_g^+)$,
$H^{34}_4(1_{\qb}^+,2_q^-,3_g^-,4_g^+)$ and $H^{34}_4(1_{\qb}^+,2_q^-,3_g^+,4_g^-)$ together
with the related coefficients that can be obtained from the base set.}
\label{table:aqgg}
\end{center}
\end{table}
\renewcommand{\baselinestretch}{1}

\subsection{Boxes}
\subsubsection{$d_{3\x21\x4}$}
\begin{eqnarray}
d_{3\x21\x4}^{(0)}(1^+_{\bar{q}},2^-_q,3^+_g,4^+_g)&=&
      -2\,\frac{\spa2.4\,\spa2.3\,\Trtxudxqxud }
   {\spa1.2\,\spa3.4^3} \nonumber \\
   &-&\frac{1}{2} \,\frac{\spb1.3\,\spb1.4\,s_{1234}}{\spb1.2\,\spa3.4}
    -\frac{1}{2}\,\frac{\spab2.(1+3).4\,\spab2.(1+4).3}{\spa1.2\,\spa3.4} \nonumber \\
   &+&2\, m^2 \Bigg[\frac{\spb1.3\,\spb1.4}{\spb1.2\,\spa3.4} 
    +3\,\frac{\spa2.3\,\spa2.4\,\spb3.4}{\spa1.2\,\spa3.4^2} \Bigg]
\end{eqnarray}
where we have introduced the notation, (c.f.~Eqs.~(\ref{tracenotation}))
\begin{equation} \label{Explicittrace}
\Trtxudxqxud=\spab3.(1+2).4 \, \spab4.(1+2).3=(s_{13}+s_{23})\,(s_{14}+s_{24})-s_{12}\,s_{34}
\end{equation}

\subsubsection{$d_{4\x3\x21}$}
\begin{eqnarray}
d_{4\x3\x21}(1^+_{\bar{q}},2^-_q,3^+_g,4^+_g)&=&
   \frac{1}{2} \, \frac{\spb3.4\,\spab2.(1+3).4^2}{\spa1.2\,\spab3.(1+2).4}
  +\frac{1}{2} \, \frac{\spb1.3^2\,\spb3.4\, s_{1234}}{\spb1.2\,\spab4.(1+2).3} \nonumber \\
  &+&2\, m^2\,\Bigg[\frac{\spa2.3\,\spb3.4\,\spab2.(1+3).4}{\spa1.2\,\spa3.4\,\spab3.(1+2).4}
    -\frac{\spb1.3\,\spb3.4\,\spab4.(2+3).1}{\spb1.2\,\spa3.4\,\spab4.(1+2).3}\Bigg]
\end{eqnarray}

\subsection{Triangles}
\subsubsection{$c_{3\x21}$}
\begin{eqnarray}
c_{3\x21}(1^+_{\bar{q}},2^-_q,3^+_g,4^+_g)&=&  2\,(s_{13}+s_{23})\,\frac{\spa2.3\,\spa2.4}{\spa1.2\,\spa3.4^3}
\end{eqnarray}

\subsubsection{$c_{12\x34}^{(0)},c_{12\x34}^{(2)}$}
\begin{eqnarray}
c_{12\x34}^{(0)}(1^+_{\bar{q}},2^-_q,3^+_g,4^+_g)&=&
   \frac{\spb1.4^2(s_{123}-s_{124}-s_{34})}{\spb1.2\,\spa3.4\,\spab3.(1+2).4}
  -\frac{\spa2.3^2\,\spb3.4^2}{\spa1.2\,\spa3.4\,\spab3.(1+2).4}\nonumber\\
  &+&\frac{\spb1.3^2(s_{124}-s_{123}-s_{34})}{\spb1.2\,\spa3.4\,\spab4.(1+2).3}
   -\frac{\spa2.4^2\,\spb3.4^2}{\spa1.2\,\spa3.4\,\spab4.(1+2).3}\nonumber\\
  &-&\frac{4\,\spb1.3\,\spb1.4}{\spb1.2\,\spa3.4}
\end{eqnarray}
\begin{eqnarray}
c_{12\x34}^{(2)}(1^+_{\bar{q}},2^-_q,3^+_g,4^+_g)&=& \frac{4}{\spa3.4^2\,\spab3.(1+2).4}
   \Bigg[\frac{\spb1.4^2\,\spa3.4}{\spb1.2}-\frac{\spa2.3^2\,\spb3.4}{\spa1.2}\Bigg] \nonumber\\
  &+&\frac{4}{\spa3.4^2\,\spab4.(1+2).3}
   \Bigg[\frac{\spb1.3^2\,\spa3.4}{\spb1.2}-\frac{\spa2.4^2\,\spb3.4}{\spa1.2}\Bigg]
\end{eqnarray}

\subsubsection{$c_{4\x123}^{(0)},c_{4\x123}^{(2)}$}
\begin{eqnarray}
c_{4\x123}^{(0)}(1^+_{\bar{q}},2^-_q,3^+_g,4^+_g)&=&
  \frac{2\,\spab2.(1+3).4^3}{\spa1.2\,\spa2.3\,\spab3.(1+2).4s_{123}}\nonumber\\
  &-&(s_{14}+s_{24}+s_{34})\,
 \Bigg[\frac{\spab2.(1+4).3\,\spab2.(1+3).4}{\spa1.2\,\spa3.4\, \Trtxudxqxud } \nonumber \\
   &+&\frac{\spb1.3\,\spb1.4\,s_{1234}}{\spb1.2\,\spa3.4\,\Trtxudxqxud }
+\frac{2\,\spa2.3\,\spa2.4}{\spa1.2\,\spa3.4^3} \Bigg]
\end{eqnarray}
\begin{eqnarray}
c_{4\x123}^{(2)}(1^+_{\bar{q}},2^-_q,3^+_g,4^+_g)&=&  \frac{4}{\spab3.(1+2).4}\Bigg\{
  \frac{(s_{14}+s_{24}+s_{34})} {\spab4.(1+2).3} 
   \Big[\frac{\spa2.3\,\spa2.4\,\spb3.4}{\spa1.2\,\spa3.4^2}+\frac{\spb1.3\,\spb1.4}{\spb1.2\,\spa3.4}\Big] \nonumber \\
   &-&2 \, \frac{\spab2.(1+3).4^3}{\spa1.2\,\spa2.3\,(s_{14}+s_{24}+s_{34})\,s_{123}} \Bigg\}
\end{eqnarray}

\subsubsection{$c_{3\x412}^{(0)},c_{3\x412}^{(2)}$}
\begin{eqnarray}
c_{3\x412}^{(0)}(1^+_{\bar{q}},2^-_q,3^+_g,4^+_g)&=&
   +2\,\frac{\spab2.(1+4).3^2\,\spab1.(2+4).3}
   {\spa2.1\,\spa1.4\,\spab4.(1+2).3\,s_{124}} \nonumber \\
   &+&(s_{13}+s_{23}+s_{34})\,\Bigg[
   -2\,\frac{\spa2.4\,\spa2.3}{\spa2.1\,\spa4.3^3}\nonumber \\
  &+&\frac{\spb1.4\,\spb1.3\,\spb4.3\,\spa4.3}
   {\spb2.1\,\spa4.3\,\Trtxudxqxud }
   +\frac{\spa2.4\,\spa2.3\,\spb4.3^2}
   {\spa2.1\,\spa4.3\,\Trtxudxqxud }\nonumber \\
   &-&\frac{\spb1.4^2}{\spb2.1\,\spa4.3\,\spab3.(1+2).4 }
   -\frac{\spb1.3^2}{\spb2.1\,\spa4.3\,\spab4.(1+2).3} \Bigg]
\end{eqnarray} 
\begin{eqnarray}
c_{3\x412}^{(2)}(1^+_{\bar{q}},2^-_q,3^+_g,4^+_g)&=&
   4\,\frac{(s_{13}+s_{23}+s_{34})}{\Trtxudxqxud}
   \Bigg[\frac{\spa2.4\,\spa2.3\,\spb4.3}{\spa2.1\,\spa4.3^2}
   +\frac{\spb1.4\,\spb1.3}{\spb2.1\,\spa4.3}\Bigg]
\nonumber \\
   &-&8\,\frac{\spab2.(1+4).3^2\,\spab1.(2+4).3}
   {\spa2.1\,\spa1.4\,\spab4.(1+2).3\,(s_{13}+s_{23}+s_{34})\,s_{124}} 
\end{eqnarray} 

\subsection{Bubbles}
\subsubsection{$b_{12}$}
\begin{eqnarray}
b_{12}(1^+_{\bar{q}},2^-_q,3^+_g,4^+_g)&=&
  \frac{4}{\spa3.4^2} \Big[ \frac{\spb1.3\,\spa2.3}{(s_{13}+s_{23})}-\frac{\spb1.4\,\spa2.4}{(s_{14}+s_{24})}\Big]
\end{eqnarray}

\subsubsection{$b_{123}$}
\begin{eqnarray}
b_{123}(1^+_{\bar{q}},2^-_q,3^+_g,4^+_g)&=& 
      \frac{4}{\spa1.2\,\spa2.3\,\spa3.4^2}
   \,\Big[\frac{\spa3.4\,\spa2.4\,\spab2.(1+3).4^2}
   {(s_{14}+s_{24}+s_{34})^2}\nonumber \\
   &-&\frac{\spa2.3\,\spa2.4\,\spab2.(1+3).4}{(s_{14}+s_{24}+s_{34})}
   -\frac{\spa1.2\,\spb1.3\,\spa2.3^2}{(s_{13}+s_{23})}\Big]
\end{eqnarray}

\subsubsection{$b_{412}$}
\begin{eqnarray}
b_{412}(1^+_{\bar{q}},2^-_q,3^+_g,4^+_g)&=& 
 4\,\frac{1}{\spa1.2\,\spa1.4\,\spa3.4^2} \,\Bigg[
 \frac{\spa2.3\,\spa3.4\,\spab2.(1+4).3\,\spab1.(2+4).3}{(s_{13}+s_{23}+s_{34})^2} \nonumber \\
 &+&\frac{\spa1.3\,\spa2.4\,\spab2.(1+4).3}{(s_{13}+s_{23}+s_{34})}
   -\frac{\spa1.2\,\spa2.4\,s_{14}}{(s_{14}+s_{24})}\Bigg]
\end{eqnarray} 

\subsubsection{$b_{1234}$}
The final bubble coefficient is given by the relation,
\begin{equation}
b_{1234}(1_{\qb}^+,2_q^-,3_g^+,4_g^+) = -b_{12}-b_{123}-b_{412} \,,
\end{equation}
where momentum and helicity labels have been suppressed on the right-hand side.

\subsection{Rational terms} 
\begin{eqnarray}
 r(1_{\qb}^+,2_q^-,3_g^+,4_g^+) &=& \frac{1}{2} \, \big[
           c_{12\x34}^{(2)}(1_{\qb}^+,2_q^-,3_g^+,4_g^+)
          +c_{4\x123}^{(2)}(1_{\qb}^+,2_q^-,3_g^+,4_g^+)\nonumber \\
 && \quad +c_{3\x412}^{(2)}(1_{\qb}^+,2_q^-,3_g^+,4_g^+)\big]
\end{eqnarray}

\section{Coefficients for $H^{34}_4(\qb^+,q^-,g^-,g^+;\Higgs)$}

The coefficients that must be computed for this amplitude are shown in
Table~\ref{table:aqgg}.

\subsection{Boxes}
\subsubsection{$d_{3\x21\x4}$}
\begin{eqnarray}
d_{3\x21\x4}^{(0)}(1^+_{\bar{q}},2^-_q,3^-_g,4^+_g)&=&
\frac{1}{2} \,\spab3.(1+2).4 \Bigg[ \frac{\spb1.4^2}{\spb2.1\,\spb4.3} -\frac{\spa2.3^2}{\spa1.2\,\spa3.4}\Bigg] \nonumber \\
  &+& 2\,m^2 \,\frac{\spab3.(1+2).4}{\spab4.(1+2).3} \,
\Bigg[\frac{\spa2.4\,\spa2.3}{\spa1.2\,\spa3.4}-\frac{\spb1.3\,\spb1.4}{\spb1.2\,\spb3.4} \Bigg]
\end{eqnarray}

\subsubsection{$d_{4\x3\x21}$}
\begin{eqnarray}
d_{4\x3\x21}^{(0)}(1^+_{\bar{q}},2^-_q,3^-_g,4^+_g)&=&
\frac{1}{2}\,\frac{s_{123}}{\spab4.(1+2).3} \, \Bigg[
\frac{\spa2.3^2\,\spb3.4}{\spa1.2}
    +\frac{\spb1.4^2\,\spa3.4}{\spb1.2}
  -4 \, \frac{\spb1.3\,\spab4.(2+3).1\, s_{34}\, s_{123}}{\spb1.2\,\spab4.(1+2).3^2}\Bigg]\nonumber\\
&-& 2 \, \frac{m^2}{\spab4.(1+2).3} \,\Bigg[\frac{3\,\spb1.3\,\spab3.(1+2).4\,\spab4.(2+3).1}{\spb1.2\,\spab4.(1+2).3}
+\frac{\spa2.3\,\spab2.(1+3).4}{\spa1.2} \Bigg]
\end{eqnarray}

\subsection{Triangles}
\subsubsection{$c_{3\x4}$}
\begin{eqnarray}
  c_{3\x4}(1^+_{\bar{q}},2^-_q,3^-_g,4^+_g) &=& \Bigg\{ \frac{2\,s_{34}\,\spb1.3\,\spa2.3}{\spab4.(1+2).3^2}
  + \frac{2\, s_{34}\,\spb1.3^2\,\spa3.4\,(2\,s_{12}+s_{13}+s_{23})}{\spb1.2\,\spab4.(1+2).3^3} \Bigg\} \nonumber \\
  &-& \Bigg\{ 1 \leftrightarrow 2, 3 \leftrightarrow 4, \langle\,\rangle\leftrightarrow[\,] \Bigg\}
\end{eqnarray}

\subsubsection{$c_{3\x21}$}
\begin{eqnarray}
c_{3\x21}(1^+_{\bar{q}},2^-_q,3^-_g,4^+_g)&=&  
2 \, (s_{13}+s_{23})\, s_{123}\, \frac{\spb1.3\,\spab4.(2+3).1}{\spb1.2\,\spab4.(1+2).3^3}
\end{eqnarray}

\subsubsection{$c_{12\x34}^{(0)},c_{12\x34}^{(2)}$}
\begin{eqnarray}
c_{12\x34}^{(0)}(1^+_{\bar{q}},2^-_q,3^-_g,4^+_g)&=&
   8\, (s_{124}-s_{123})\, (s_{12}+s_{34}+2\,s_{13}+2\,s_{23})\,\frac{\spa2.4\,\spb1.3\,\spab3.(1+2).4}
{\spab4.(1+2).3^2 \DeltaThree(1,2,3,4)}\nonumber\\
  &+&\Big((9\,s_{13}-7\,s_{23}-s_{14}-s_{24}+4\,s_{34})\,\spa2.4\,\spb1.4 \nonumber \\
         &-&(9\,s_{14}-7\,s_{24}-s_{13}-s_{23}+4\,s_{34})\,\spa2.3\,\spb1.3\Big)\times\frac{1}{\spab4.(1+2).3^2}\nonumber\\
  &+&12\,\frac{s_{1234}\,((s_{13}+s_{23})^2-(s_{14}+s_{24})^2)\,\spab2.(3+4).1\,\spab3.(1+2).4}
     {\spab4.(1+2).3 \DeltaThree(1,2,3,4)^2}\nonumber\\
  &+&4\,
 \Big(\big\{3\,(s_{12}+s_{34})+4\,(s_{13}+s_{23}+s_{14})\big\} \, \spb1.3\,\spa2.3 \nonumber \\
   &-&\big\{3\,(s_{12}+s_{34})+4\,(s_{13}+s_{24}+s_{14})\big\} \, \spb1.4\,\spa2.4 \Big)
 \times \frac{\spab3.(1+2).4 }{\spab4.(1+2).3\,\DeltaThree(1,2,3,4)}
\nonumber \\
  &-&24\, \frac{\spb1.3\,\spa2.4\,\spab3.(1+2).4^2}{\spab4.(1+2).3\,\DeltaThree(1,2,3,4)}
  -8\,\frac{\spb1.4\,\spa2.3\,\spab3.(1+2).4}{\DeltaThree(1,2,3,4)}
  +8\,\frac{\spb1.4\,\spa2.3}{\spab4.(1+2).3} \nonumber \\
  &+&\symbrack\Bigg\{\frac{2\,\spa2.4^2\,\spb3.4\,(s_{14}+s_{24})^2}{\spa1.2\,\spab4.(1+2).3^3}
  +\frac{\spb1.3\,\spa2.4 \,(s_{14}+s_{24})\,(4\,s_{124}-2\,s_{34})}{\spab4.(1+2).3^3}\nonumber\\
  &+&\frac{(s_{13}+s_{23})\,\spa2.3\,\spa2.4\,(s_{14}+s_{24}-s_{13}-s_{23})}{\spa1.2\,\spa3.4\,\spab4.(1+2).3^2}
  +\frac{2\,\spa2.3\,\spa2.4\,\spb3.4\,(s_{14}+s_{24})}{\spa1.2\,\spab4.(1+2).3^2}\symbrack\Bigg\}\nonumber\\
  &-&\Bigg\{1 \leftrightarrow 2, 3 \leftrightarrow 4, \langle\,\rangle\leftrightarrow[\,] \Bigg\} 
\end{eqnarray}
\begin{eqnarray}
c_{12\x34}^{(2)}(1^+_{\bar{q}},2^-_q,3^-_g,4^+_g) &=&
\Big\{\frac{4\,\spa2.3^2}{\spa1.2\,\spa3.4\,\spab4.(1+2).3}
-\frac{4\,\spa2.4^2\,\spab3.(1+2).4}{\spa1.2\,\spa3.4\,\spab4.(1+2).3^2}\nonumber\\
  &-& 8\,(s_{13}+s_{23}+s_{14}+s_{24})\frac{\spa2.3\,\spa2.4\,\spab3.(1+2).4}
 {\spa1.2\,\spa3.4\,\spab4.(1+2).3\,\DeltaThree(1,2,3,4)} \nonumber \\
  &-&16\,\frac{\spb1.3\,\spa2.3\,\spab3.(1+2).4}{\spab4.(1+2).3 \,\DeltaThree(1,2,3,4)}\Big\}\nonumber\\
  &-&\Big\{ 1 \leftrightarrow 2, 3 \leftrightarrow 4, \langle\,\rangle\leftrightarrow[\,] \Big\}
\end{eqnarray}

\subsubsection{$c_{4\x123}^{(0)},c_{4\x123}^{(2)}$}
\begin{eqnarray}
c_{4\x123}^{(0)}(1^+_{\bar{q}},2^-_q,3^-_g,4^+_g)&=&
 -\frac{2\,(s_{14}+s_{24}+s_{34})\,\spb1.3\,\spab4.(2+3).1\,s_{123}}{\spb1.2\,\spab4.(1+2).3^3}\nonumber \\
  &-&\frac{\spa2.3^2\,(s_{14}+s_{24}+s_{34})}{\spa1.2\,\spa3.4\,\spab4.(1+2).3}
  -\frac{\spb1.4^2\,(s_{14}+s_{24}+s_{34})}{\spb1.2\,\spb3.4\,\spab4.(1+2).3}
  +\frac{2\,\spb1.4^2\,\spb2.4}{\spb1.2\,\spb2.3\,\spb3.4}
\end{eqnarray}
\begin{eqnarray}
c_{4\x123}^{(2)}(1^+_{\bar{q}},2^-_q,3^-_g,4^+_g)&=&
 \frac{4\,(s_{14}+s_{24}+s_{34})}{\spab4.(1+2).3^2}
 \Big[\frac{\spa2.3\,\spa2.4}{\spa1.2\,\spa3.4}+\frac{\spb1.3\,\spb1.4}{\spb1.2\,\spb3.4}\Big]\nonumber \\
  &-&8\,\frac{\spb1.4^2\,\spb2.4}{\spb1.2\,\spb2.3\,\spb3.4\,(s_{14}+s_{24}+s_{34})}
\end{eqnarray}

\subsection{Bubbles}
\subsubsection{$b_{34}$}
\begin{eqnarray}
b_{34}(1^+_{\bar{q}},2^-_q,3^-_g,4^+_g)&=&
   2\,\frac{\spab3.(1+2).4}{s_{12}\,\spab4.(1+2).3\,\DeltaThree(1,2,3,4)}
   \,\Big[3\,\frac{
   \spab2.(3+4).1\,(s_{124}^2-s_{123}^2)\,(s_{13}+s_{14}+s_{23}+s_{24})}{\DeltaThree(1,2,3,4)} \nonumber \\
    &+&2\,\frac{(s_{124}-s_{123})\,(\spa1.2\,\spb1.3\,\spab4.(2+3).1-\spa2.4\,\spb1.2\,\spab2.(1+4).3)}{\spab4.(1+2).3}
   \nonumber \\
    &-&3\,(s_{123}+s_{124})\,(\spa2.3\,\spb1.3-\spa2.4\,\spb1.4) \Big]
\end{eqnarray}

\subsubsection{$b_{12}$}
\begin{eqnarray}
b_{12}(1^+_{\bar{q}},2^-_q,3^-_g,4^+_g)&=&
\Bigg\{\frac{4}{\spab4.(1+2).3^2}
   \,\Bigg[
   -\frac{\spa1.2\,\spb1.3^2\,\spab3.(1+2).4}{\spb3.4\,(s_{13}+s_{23})} \nonumber \\
   &+&\frac{\spab3.(1+2).4}{\spa3.4\,\DeltaThree(1,2,3,4)} \,\Big[
   s_{12}\, \spa2.4\,\spb1.3\,\spa3.4 
   -2\,\spa1.2\,\spb1.3^2\,\spa3.4^2 \nonumber \\
   &&-\spa2.3\,\spab4.(1+2).3\,\spab4.(2+3).1
     -\spa2.4^2\,\spb1.2\,(s_{13}+s_{23}+s_{14}+s_{24}) \Big] \Bigg]\nonumber \\
   &-&12\,(s_{13}+s_{23})\,(2\,s_{12}+s_{13}+s_{23})\,\frac{\spab2.(3+4).1\,\spab3.(1+2).4}
  {\spab4.(1+2).3\,\DeltaThree(1,2,3,4)^2}\Bigg\} \nonumber \\
&-& \Bigg\{1 \leftrightarrow 2, 3 \leftrightarrow 4, \langle\,\rangle\leftrightarrow[\,] \Bigg\}
\end{eqnarray}

\subsubsection{$b_{123}$}
\begin{eqnarray}
b_{123}(1^+_{\bar{q}},2^-_q,3^-_g,4^+_g)&=& 
      4\,s_{123}\,\Big[
   \frac{\spb1.4^2\,\spab4.(1+3).2}
   {\spb1.2\,\spb2.3\,\spab4.(1+2).3\,(s_{14}+s_{24}+s_{34})^2}\nonumber \\
   &+&\frac{\spb1.4\,\spab4.(2+3).1}
   {\spb1.2\,\spab4.(1+2).3^2\,(s_{14}+s_{24}+s_{34})}
   +\frac{\spa2.3\,\spb1.3}
    {(s_{13}+s_{23})\,\spab4.(1+2).3^2}\Big]
\end{eqnarray}

\subsubsection{$b_{1234}$}
The final bubble coefficient is given by the relation,
\begin{equation}
b_{1234}(1_{\qb}^+,2_q^-,3_g^-,4_g^+) = -b_{12}-b_{34}-b_{123}-b_{412} \,,
\end{equation}
where momentum and helicity labels have been suppressed on the right-hand side.

\subsection{Rational terms} 
\begin{eqnarray}
 r(1_{\qb}^+,2_q^-,3_g^-,4_g^+) &=& \frac{1}{2} \, \big[
           c_{12\x34}^{(2)}(1_{\qb}^+,2_q^-,3_g^-,4_g^+)
          +c_{4\x123}^{(2)}(1_{\qb}^+,2_q^-,3_g^-,4_g^+)\nonumber \\
 && \quad -c_{4\x123}^{(2)}(2_{\qb}^+,1_q^-,4_g^-,3_g^+)|_{[\,]\leftrightarrow \langle\,\rangle}\big]
\end{eqnarray}

\section{Coefficients for $H^{34}_4(\qb^+,q^-,g^+,g^-;\Higgs)$}
Most of the coefficients for this amplitude can be easily obtained from those for
$H^{34}_4(\qb^+,q^-,g^-,g^+)$ by performing the following operation:
\begin{equation}
1 \leftrightarrow 2 \,, \quad \langle\,\rangle \leftrightarrow [\,] \,.
\end{equation}
However, for some coefficients, this procedure effectively changes the colour ordering of the
gluons in the sub-amplitude.  For this reason it is necessary to specify the four coefficients
shown in Table~\ref{table:aqgg}. Results for the base set of coefficients are given explicitly here.

\subsection{Triangle}
\subsubsection{$c_{4\x123}^{(0)},c_{4\x123}^{(2)}$}
\begin{eqnarray}
c_{4\x123}^{(0)}(1^+_{\bar{q}},2^-_q,3^+_g,4^-_g)&=&
   -2\,\frac{\spa3.4\,\spab2.(1+3).4^2\,s_{123}}{\spa1.2\,\spab3.(1+2).4^3} 
   +2\,\frac{\spa2.4\,\spab2.(1+3).4\,s_{123}}{\spa1.2\,\spab3.(1+2).4^2} \nonumber \\
 &-&\frac{\spa2.4^2\,\spab2.(1+3).4}{\spa1.2\,\spa2.3\,\spab3.(1+2).4} 
  +\frac{\spb1.3^2\,(s_{14}+s_{24}+s_{34})}{\spb1.2\,\spb3.4\,\spab3.(1+2).4}\nonumber \\
  &-&\frac{\spa2.4^3}{\spa1.2\,\spa2.3\,\spa3.4}
\end{eqnarray}
\begin{eqnarray}
c_{4\x123}^{(2)}(1^+_{\bar{q}},2^-_q,3^+_g,4^-_g)&=& -4\frac{(s_{14}+s_{24}+s_{34})}{\spab3.(1+2).4^2}
   \Bigg[\frac{\spa2.3\,\spa2.4}{\spa1.2\,\spa3.4}
   +\frac{\spb1.3\,\spb1.4}{\spb1.2\,\spb3.4}\Bigg] \nonumber  \\
   &+&8\,\frac{\spa2.4^3}{\spa1.2\,\spa2.3\,\spa3.4\,(s_{14}+s_{24}+s_{34})}
\end{eqnarray}

\subsection{Bubble}
\subsubsection{$b_{123}$}
\begin{eqnarray}
b_{123}(1^+_{\bar{q}},2^-_q,3^+_g,4^-_g)&=& 
      4\,s_{123}\,\Big[
   -\frac{\spa2.4^2\,\spab2.(1+3).4}
   {\spa1.2\,\spa2.3\,\spab3.(1+2).4\,(s_{14}+s_{24}+s_{34})^2} \nonumber \\
    &-&\frac{\spa2.4\,\spab2.(1+3).4}
   {\spa1.2\,\spab3.(1+2).4^2\,(s_{14}+s_{24}+s_{34})}
   +\frac{\spb1.3\,\spa2.3}
   {(s_{13}+s_{23})\,\spab3.(1+2).4^2}\Big]
\end{eqnarray}

\subsection{Rational terms} 
\begin{eqnarray}
 r(1_{\qb}^+,2_q^-,3_g^+,4_g^-) &=& \frac{1}{2} \, \big[
           c_{12\x34}^{(2)}(2_{\qb}^+,1_q^-,3_g^-,4_g^+)|_{[\,]\leftrightarrow \langle\,\rangle}
          +c_{4\x123}^{(2)}(1_{\qb}^+,2_q^-,3_g^+,4_g^-)\nonumber \\
 && \quad -c_{4\x123}^{(2)}(2_{\qb}^+,1_q^-,4_g^+,3_g^-)|_{[\,]\leftrightarrow \langle\,\rangle}\big]
\end{eqnarray}

\section{Amplitude for $0 \to {\qb}q{\qb}q \Higgs $}

The amplitude for the $gg\Higgs$ process with off-shell gluons with momenta $k_1,k_2$ has been given in ref.~\cite{DelDuca:2001fn}.
Thus our result is exactly given by ref.~\cite{DelDuca:2001fn} and is only included here for completeness.

The amplitude can be obtained by considering the tensor current for $0 \to gg\Higgs$ with two off-shell gluons
(with momenta $k_1$ and $k_2$),
\begin{equation}
{\cal T}^{\mu_1\mu_2}(k_1,k_2) = -i \delta^{c_1 c_2} \, \frac{g_s^2}{8 \pi^2} \, \Big(\frac{m^2}{v}\Big) \; \Big[
F_T(k_1,k_2) \, T_T^{\mu_1\mu_2}+ F_L(k_1,k_2) \, T_L^{\mu_1\mu_2}\Big] \,.
\end{equation}
The two tensor structures appearing here are,
\begin{eqnarray}
T_T^{\mu_1\mu_2} &=&k_1 \cdot k_2 \, g^{\mu_1 \mu_2}-k_1^{\mu_2} k_2^{\mu_1} \\
T_L^{\mu_1\mu_2} &=&k_1^2 k_2^2 \, g^{\mu_1 \mu_2}
-k_1^2 \, k_2^{\mu_1} k_2^{\mu_2} 
-k_2^2 \, k_1^{\mu_1} k_1^{\mu_2} 
+k_1\cdot k_2 \, k_1^{\mu_1} k_2^{\mu_2} 
\label{4qtensors}
\end{eqnarray}
and the form factors are given by\footnote{Note that to produce our standard overall normalization for the 
helicity amplitude we have changed the normalization of the form factors by a factor of 2 with respect to ref.~\cite{DelDuca:2001fn}.}
\begin{eqnarray}
\label{FLequation}
\label{FTequation}
F_T(k_1,k_2) &=& 
-\frac{1}{\Delta(k_1,k_2)} 
    \Big\{k_{12}^2 \,(B_0(k_1;m)+B_0(k_2;m)-2 B_0(k_{12};m)-2 k_1\cdot k_2 \, C_0(k_1,k_2;m)) \nonumber \\
&+& (k_1^2-k_2^2)\,(B_0(k_1;m)-B_0(k_2;m))\Big\}-\, k_1 \cdot k_2 \, F_L(k_1,k_2)  \\
F_L(k_1,k_2) &=& 
-\frac{1}{\Delta(k_1,k_2)} 
    \Big\{
    \big[2-\frac{3 k_1^2 \, k_2\cdot k_{12}}{\Delta(k_1,k_2)}\big]\,(B_0(k_1;m)-B_0(k_{12};m)) \nonumber \\
&+& \big[2-\frac{3 k_2^2 \, k_1\cdot k_{12}}{\Delta(k_1,k_2)}\big]\,(B_0(k_2;m)-B_0(k_{12};m)) \nonumber \\
&-&\Big[4 m^2 +k_1^2+k_2^2 +k_{12}^2 -3 \frac{k_1^2 \,k_2^2 \, k_{12}^2}{\Delta(k_1,k_2)}\Big]\, C_0(k_1,k_2;m)-2\Big\}
\end{eqnarray}
where $k_{12}=k_1+k_2$ and $\Delta(k_1,k_2)=k_1^2 \, k_2^2 -(k_1 \cdot k_2)^2$.
By contracting Eq.~(\ref{4qtensors}) with currents for the quark-antiquark lines we then arrive at the result for
the amplitude.  All helicity combinations can be obtained from permutations of the single expression,
\begin{eqnarray}
H^{4q}_{4}(1^{+}_{{\qb}},2^{-}_{q},3^{+}_{{\qb}^\prime},4^{-}_{q^\prime};\Higgs)
&=&
    \Big[\frac{\spab2.(3+4).1\,\spab4.(1+2).3+\spa2.4\,\spb1.3\,(2\, p_{12}.p_{34})}{s_{12}\,s_{34}}\Big]\, F_T(p_{12},p_{34}) \nonumber \\
          &+& 2\, \spa2.4 \,\spb1.3\, F_L(p_{12},p_{34}) \,.
\end{eqnarray}

\section{Large mass limit}
In the limit of large (top quark) mass our results agree with those obtained in the effective field theory
given by Eq.~(\ref{EFT}).  For $gggg \Higgs$ we have~\cite{Dawson:1991au,Kauffman:1996ix,Badger:2009hw},
\begin{eqnarray}
m^2 H_4^{1234}(1^+,2^+,3^+,4^+;\Higgs) &\to & \frac{2}{3} \frac{s_{1234}^2}{\spa1.2\spa2.3\spa3.4\spa4.1}\\
m^2 H_4^{1234}(1^+,2^+,3^+,4^-;\Higgs) &\to& \frac{2}{3} \Big[
      -\frac{s_{1234}^2\,\spb1.3^4}
      {s_{134}\,\spb4.3\,\spb4.1\,\spab2.(1+4).3\,\spab2.(3+4).1} \nonumber \\
      &+&\frac{\spab4.(2+3).1^3}
      {s_{234}\,\spab2.(3+4).1\,\spa3.4\,\spa2.3}
      +\frac{\spab4.(2+1).3^3}
      {s_{124}\,\spab2.(1+4).3\,\spa1.4\,\spa2.1}\Big]\\
\label{EFTppmm}
m^2 H_4^{1234}(1^+,2^+,3^-,4^-;\Higgs) &\to & \frac{2}{3} 
    \Big[\frac{\spb1.2^4}{\spb1.2\,\spb2.3\,\spb3.4\,\spb4.1}
        +\frac{\spa3.4^4}{\spa1.2\,\spa2.3\,\spa3.4\,\spa4.1}\Big]\, .
\end{eqnarray}
In this limit the amplitudes satisfy the dual Ward identity. For example,
\begin{equation}
  H_4^{1234}(1^+,2^-,3^+,4^-;\Higgs) 
 +H_4^{1234}(3^+,1^+,2^-,4^-;\Higgs)
 +H_4^{1234}(1^+,3^+,2^-,4^-;\Higgs)=0\,,
\end{equation}
which means that in this limit our amplitude $H_4^{1234}(1^+,2^-,3^+,4^-;\Higgs)$ is related
to two amplitudes of the form already specified in Eq.~(\ref{EFTppmm}).
We note in passing that the dual Ward identity means that, in the effective theory, the
subleading-color term represented by the second line of Eq.~(\ref{coloursum}) is absent.

For the large mass limit of the $\qb q gg \Higgs$ amplitudes we have~\cite{Kauffman:1996ix},
\begin{eqnarray}
     m^2 \, H^{34}(1_{\qb}^+,2_q^-,3_g^+,4_g^+;\Higgs)&\to&
 \frac{2}{3}\, \Bigg[\frac{\spab2.(1+4).3^2\,\spb1.4}{s_{124}\,\spa2.4}
   \,\Big[\frac{1}{s_{12}}+\frac{1}{s_{14}}\Big] \nonumber \\
   && \quad -\frac{\spab2.(1+3).4^2 \spb1.3}{s_{123}\, s_{12} \spa2.3} 
   +\frac{\spab2.(3+4).1^2}{\spb1.2 \, \spa2.3\, \spa2.4\,\spa3.4}\Bigg] \\
m^2 \, H^{34}(1_{\qb}^+,2_q^-,3_g^+,4_g^-;\Higgs) &\to&
   \frac{2}{3}\, \Bigg[  \frac{\spa2.4^3}{\spa1.2\,\spa2.3\,\spa3.4}
                        -\frac{\spb1.3^3}{\spb1.2\,\spb1.4\,\spb3.4}\Bigg] \\
m^2 \,H^{34}(1_{\qb}^+,2_q^-,3_g^-,4_g^+;\Higgs)  &\to&
   \frac{2}{3}\, \Bigg[  \frac{\spa2.3^2\,\spa1.3}{\spa1.2\,\spa1.4\,\spa3.4}
                        -\frac{\spb1.4^2\,\spb2.4}{\spb1.2\,\spb2.3\,\spb3.4}\Bigg]
\end{eqnarray}
and for $\qb q \qb^\prime q^\prime$,
\begin{equation}
m^2 \, H^{4q}(1_{\qb}^+,2_q^-,3_{\qb^\prime}^+,4_{q^\prime}^-;\Higgs) \to
    -\frac{2}{3}\, \Bigg[\frac{\spb1.3^2}{\spb1.2\,\spb3.4} +\frac{\spa2.4^2}{\spa1.2\,\spa3.4}\Bigg] \,.
\end{equation}

\section{Conclusions}
We have presented analytic results for all helicity amplitudes
representing the processes $0 \to g g g g \Higgs$, $0 \to \qb q gg \Higgs$ and
$0\to \qb q \qb^\prime q^\prime\Higgs$, where the interaction is mediated by a loop of
massive fermions and all dependence on the fermion mass is retained.
In order to obtain compact results we have used unitarity techniques
and also exploited the correspondence between this amplitude and the
one in which the massive fermion is replaced by a coloured scalar.  In
order to further simplify our analytic results to the forms presented
here we have supplemented this approach by the use of momentum
twistors and reconstruction from high-precision numerical evaluations.
In combination this powerful set of tools rendered this calculation
tractable.

Our results for the amplitudes were checked using an in-house implementation of
the $D$-dimensional unitarity method~\cite{Ellis:2008ir} and also against a previous
unitarity-based calculation~\cite{Neumann:2016dny}.  Complete agreement was found at the 
amplitude level.  Our results for the squared matrix elements are also in full agreement
with those obtained using the code OpenLoops 2~\cite{Buccioni:2019sur}.
A comparison of the evaluation time of squared matrix
elements against both the previous code implemented in MCFM~\cite{Campbell:2011bn,Campbell:2015qma,Campbell:2019dru}
and OpenLoops 2 indicates a speed-up by
at least an order of magnitude over previously-available results. 
Our results have been implemented in version 9.1 of the code MCFM\footnote{
The code can be downloaded from {\tt https://mcfm.fnal.gov}.}, that includes a
calculation of the full matrix elements for the three partonic processes under consideration.
Results are given in the subdirectory {\tt MCFM-9.1/src/ggHgg\_mass} of that release and 
the result for a particular equation can be found by searching for the
text ``Implementation'';  every file that gives the result for an integral coefficient 
contains a comment of the form:
\begin{verbatim}
         !     Implementation of Eq.~(x.xx) from arXiv:2002.04018 v2
\end{verbatim}

The results of this paper will be useful for improving calculations
of the $\Higgs+$jet process at NLO in the full theory.  The analytic forms presented here will
also be useful in their own right, for understanding the structure of gauge-theory amplitudes.
In particular the simplification of our results due to the choice of effective
pentagon integral coefficients may have deeper origins that remain to be explored.

\section*{Acknowledgements}
We would like to acknowledge useful discussions with Simon Badger and thank
Daniel Ma\^itre for the use of his code to generate spinor helicity ans\"atze.
This document was prepared using the resources of
the Fermi National Accelerator Laboratory (Fermilab), a
U.S. Department of Energy, Office of Science, HEP User
Facility. Fermilab is managed by Fermi Research Alliance, LLC (FRA),
acting under Contract No.\ DE-AC02-07CH11359.
\appendix

\section{Integrals}
\label{Integrals}
We define the denominators of the integrals as follows,
\begin{equation} \label{denominatordef}
D(\ell) = \ell^2-m^2+i\varepsilon \, .
\end{equation}
The momenta running through the propagators are,
\begin{eqnarray} \label{denominators}
\ell_1 &=& \ell+p_1 = \ell +q_1 \, \nonumber \\
\ell_{12} &=& \ell+p_1+p_2 = \ell +q_2 \, \nonumber \\
\ell_{123}&=& \ell+p_1+p_2+p_3 = \ell +q_3 \, \nonumber \\
\ell_{1234} &=& \ell+p_1+p_2+p_3+p_4 = \ell +q_4 \, .
\end{eqnarray}
The $p_i$ are the external momenta, whereas the $q_i$ are the off-set momenta in the propagators.
In terms of these denominators the integrals are,
\begin{eqnarray}
\label{Integral_defns}
B_0(p_1;m) &=& \frac{\bar\mu^{4-n}}{r_\Gamma}\frac{1}{i \pi^{n/2}} \int {\rm d}^n\ell \,\frac{1}{D(\ell)\,D(\ell_1)}\, , \nonumber \\
C_0(p_1,p_2;m) &=& \frac{1}{i \pi^{2}} \int {\rm d}^4\ell \,\frac{1}{D(\ell)\,D(\ell_1)\,D(\ell_{12})}\, ,\nonumber \\
D(\ell)(p_1,p_2,p_3;m) &=& \frac{1}{i \pi^{2}} \int {\rm d}^4\ell \,\frac{1}{D(\ell)\,D(\ell_1)\,D(\ell_{12})\,D(\ell_{123})}\, ,\nonumber \\
E_0(p_1,p_2,p_3,p_4;m) &=& \frac{1}{i \pi^{2}} \int {\rm d}^4\ell \,\frac{1}{D(\ell) D(\ell_1)\,D(\ell_{12})\,D(\ell_{123})\,D(\ell_{1234})}\, .
\end{eqnarray}
where $r_\Gamma=1/\Gamma(1-\epsilon)+O(\epsilon^3)$ and $\bar\mu$ is an arbitrary mass scale.
\section{Spinor algebra}
\label{spinorsection}
All results are presented using the
standard notation for the kinematic invariants of the process,
\begin{equation}
s_{ij} = (p_i+p_j)^2 \, ,
s_{ijk} = (p_i+p_j+p_k)^2 \, ,
s_{ijkl} = (p_i+p_j+p_k+p_l)^2 \,.
\end{equation}
and the Gram determinant,
\begin{equation} \label{Delta3eqn}
\DeltaThree(i,j,k,l) =(s_{ijkl}-s_{ij}-s_{kl})^2-4 s_{ij} s_{kl}  \, .
\end{equation}
We express the amplitudes in terms of spinor products defined as,
\begin{equation}
\label{Spinor_products1}
\spa i.j=\bar{u}_-(p_i) u_+(p_j), \;\;\;
\spb i.j=\bar{u}_+(p_i) u_-(p_j), \;\;\;
\spa i.j \spb j.i = 2 p_i \cdot p_j,\;\;\;
\end{equation}
and we further define the spinor sandwiches for massless momenta $j$ and $k$,
\begin{eqnarray}
\label{Spinor_products2}
\spab{i}.{(j+k)}.{l} &=& \spa{i}.{j} \spb{j}.{l} +\spa{i}.{k} \spb{k}.{l} \nonumber \\
\spba{i}.{(j+k)}.{l} &=& \spb{i}.{j} \spa{j}.{l} +\spb{i}.{k} \spa{k}.{l}
\end{eqnarray}
In the Weyl representation for the Dirac gamma matrices we have, 
\begin{equation} 
\slsh{p} = \gamma^0 \, p_0-\gamma^1 \, p_1-\gamma^2 \, p_2-\gamma^3 \, p_3
= \left(\begin{matrix}
0& 0 & p^+ & p^1-ip^2\cr
0& 0 & p^1+i p^2 & p^-\cr
p^- & -p^1+ip^2& 0 & 0 \cr
-p^1-i p^2 & p^+ & 0 & 0 
\end{matrix}
\right)\ ,
\end{equation}
where $p^\pm\ =\ p^0 \pm p^3$. 
The spinor solutions of the massless Dirac equation are,
\begin{equation} \label{eq:explicitspinor}
u_+(p) =
  \left[ \begin{matrix} \sqrt{p^+} \cr 
   \sqrt{p^-} e^{i\varphi_p} \cr
                  0 \cr 
   0 \cr  \end{matrix}\right] , \hskip3mm
u_-(p) =
  \left[ \begin{matrix} 0 \cr 
                  0 \cr
                \sqrt{p^-} e^{-i\varphi_p} \cr 
                                 -\sqrt{p^+} \cr  \end{matrix}\right] , 
\end{equation}
where 
\begin{equation} \label{eq:phasekdef}
e^{\pm i\varphi_p}\ \equiv\ 
  \frac{ p^1 \pm ip^2 }{ \sqrt{(p^1)^2+(p^2)^2} }
\ =\  \frac{ p^1 \pm ip^2 }{ \sqrt{p^+p^-} }\ ,
\qquad p^\pm\ =\ p^0 \pm p^3.  
\end{equation}
In this representation the Dirac conjugate spinors are,
\begin{equation}
\label{eq:explicitspinorconjg}
\bar{u}_{+}(p) \equiv  u_{+}^\dagger(p) \gamma^0 =
  \left[ 0, 0, \begin{matrix} \sqrt{p^+} , 
   \sqrt{p^-} e^{-i\varphi_p}  \cr  \end{matrix}\right] 
\end{equation}
\begin{equation}
\bar{u}_{-}(p) \equiv  u_{-}^\dagger(p) \gamma^0 =
  \left[ \sqrt{p^-} e^{i\varphi_p}, -\sqrt{p^+}, 0,0 \right] 
\end{equation}

\section{Results for tree-level amplitudes with massive scalars}
\label{scalarsection}
In the following we shall give results for colour-ordered tree amplitudes
for a scalar, anti-scalar pair coupled to $n$ partons ($n$ gluons or $n-2$ gluons with a quark antiquark pair). 
For the $n$ gluon case the amplitudes are defined as follows,
\begin{equation}
 \cA_n^\tree(\ell; 1,\ldots,n;\ellb) =  i g_s^n \sum_{\sigma\in \Sigma_{n}}
   (t^{c_{\sigma(1)}} \ldots t^{c_{\sigma(n)}})_{j\jb} \,
    A_n^\tree(\ell;\sigma(1),\ldots,\sigma(n);\ellb) \, ,
\end{equation}
where $\Sigma_{n}$ is the permutation group on $n$ elements, and
$A_n^\tree$ are the tree-level partial amplitudes. The $t$ matrices
are the SU(3) matrices in the fundamental representation normalized as
in Eq.~(\ref{normalization}).  The massive scalars are in the
fundamental representation of SU(3) and the colour indices of the
scalar and anti-scalar are $j$ and $\jb$ respectively.

We adopt throughout the convention that all momenta are taken to be
outgoing.  Correspondingly we have the tree amplitudes with an
additional Higgs boson, derived using the Lagrangian,
Eq.~(\ref{ScalarLagrangian}),
\begin{equation}
 \cA_n^\tree(\Higgs; \ell ; 1,\ldots,n;\ellb) =  i \left(\frac{-\lambda}{4}\right) g_s^n \sum_{\sigma\in \Sigma_{n}}
   (t^{a_{\sigma(1)}} \ldots t^{a_{\sigma(n)}})_{j\jb} \,
    A_n^\tree(\Higgs; \sigma(1),\ldots,\sigma(n)) \, .
\end{equation}
We define the denominators of the scalar propagators as $D(\ell) = \ell^2-m^2$,
which must be supplemented by the $+i \varepsilon$ prescription when the propagators
are used in loop diagrams, c.f.~Eq.~(\ref{denominatordef}).  
The momenta in the propagators are defined in Eq.~(\ref{denominators}), with 
similar expressions for $\ellb$.
Because the external momentum $p_1$ is light-like, for an on-shell $\ell$
we may also write $D(\ell_1)=\spab1.\ell.1$.

\subsection{One gluon}
\begin{equation} \label{onegluon}
 A^{\tree}_1(\ell;1^+;\ellb )=\frac{\spab{b}.\ell.1}{\spa{b}.1},\;\;\;\; A^{\tree}_1(\ell;1^-;,\ellb)=\frac{\spab1.\ell.b}{\spb1.b} 
\end{equation}

\begin{eqnarray} \label{onegluonplushiggs}
  A_1^\tree(\Higgs,\ell;1^+;\ellb) &=& 
 -4 \Big[\frac{\spab{b}.\ellb.1}{\spa{b}.1\,\spab1.\ellb.1} -\frac{\spab{b}.\ell.1}{\spa{b}.1 \,\spab1.\ell.1} \Big]\nonumber \\ 
&=&-4 \Big[\frac{\spab{b}.{\ellb}.1\,\spab1.\ell.1 - \spab{b}.\ell.1 \, \spab1.\ellb.1}{\spa{b}.1\,\spab1.\ellb.1 \,\spab1.\ell.1}\Big]
\end{eqnarray}
In Eqs.~(\ref{onegluon},\ref{onegluonplushiggs}) $b$ is an arbitrary light-like momentum.
\subsection{Two gluons}
\begin{eqnarray}
A^{\tree}_2(\ell;1^+,2^+;\ellb)=-\frac{m^2\,\spb1.2}{\spa1.2\,D(\ell_1)} 
\end{eqnarray}
\begin{eqnarray}
A^\tree_2(\ell;1^+,2^-;\ellb)=-\frac{\spab2.\ell.1^2}{\spab1.2.1\,D(\ell_1)} 
\end{eqnarray}

\begin{eqnarray} \!\!\!\!
A^\tree_2(\Higgs,\ell;1^+,2^+;\ellb) &=& -4 \frac{\spb1.2}{\spa1.2}
\Bigg\{\Big[\frac{m^2}{D(\ell_1) \, D(\ell_{12})}+\frac{m^2}{D(\ellb_2)\,D(\ellb_{12})}\Big] 
-\frac{1}{s_{12}} 
\Big[1-\frac{\spab1.\ellb.2\,\spab2.\ell.1}{D(\ell_1)\,D(\ellb_2)}  \Big]\Bigg\} \\  \!\!\!\!
A^\tree_2(\Higgs,\ell;1^+,2^-;\ellb) &=& -4 \frac{1}{s_{12}}
\Bigg\{     \frac{\spab2.\ell.1^2} {D(\ell_{12}) \,D(\ell_1)}
           +\frac{\spab2.\ellb.1^2}{D(\ellb_{12})\, D(\ellb_2)} 
-\frac{\spab2.\ellb.1 \spab2.\ell.1 }{D(\ell_1)\, D(\ellb_2)}\Bigg\} 
\end{eqnarray}

\subsection{Three gluons}
The spurious poles in the original BCFW form of the amplitudes~\cite{Badger:2005zh}
can be eliminated and the amplitudes rewritten in the following form~\cite{Britto:2006fc},
\begin{eqnarray}  \label{Sppm}
A^{\tree}_3(\ell;1^+,2^+, 3^-;\ellb) & = &
\frac{\spab3.{\ellb (1+2) \ell}.1\, \spab3.\ellb.2 }{\spa1.2\,s_{23}\,D(\ell_1)\,D(\ell_{12})}
+\frac{\spb1.2\,\spaa3.(1+2).\ellb.3}{\spa1.2\,s_{23} \, s_{123}}, \\
A^{\tree}_3(\ell;1^+,2^-,3^+;\ellb) & = & \frac{1}{s_{12}\, s_{23}}
 \Big[
\frac{\spb2.1 \spab2.\ell.1 \spab2.\ellb.3^2} {D(\ell) D(\ell_{12})}
+\frac{\spb3.1\,\spab2.\ell.1\, \spab2.\ellb.3 }{D(\ell_{12})} \nonumber \\
&-&\frac{\spaa2.\ellb.(1+3).2\,\spb1.3^2}{s_{123}}
 \Big].
\end{eqnarray}
By using charge conjugation on Eq.~(\ref{Sppm}) we also obtain the following result,  
\begin{eqnarray}
A^{\tree}_3(\ell;1^+,2^-, 3^-;\ellb) & = &
\frac{\spab3.{\ellb (2+3) \ell}.1 \,\spab2.{\ell}.1}{\spb2.3\,s_{12}\,D(\ell_1)\,D(\ell_{12})}
-\frac{\spa2.3\,\spbb1.(2+3).\ell.1}{\spb2.3\,s_{12}\,s_{123}}, 
\end{eqnarray}

\subsection{One gluon, two quarks}
For the calculation of the $H^{34}({\qb} q gg;\Higgs)$ we also need the tree amplitude for a pair of 
SU(3) triplet scalars coupled to a quark-antiquark pair with an additional gluon.  
\begin{eqnarray}
 {\cA}^{{\qb}qg}_3(\ell;1_{\qb}^+,2_{q}^-, 3^+_g;\ellb) & = &
  i \sqrt{2} g_s^3\, \Bigg\{ 
 \Big[(t^{c_3})_{j j_1}\,\delta_{j_2,\jb}-(t^{c_3})_{j_2 \jb}\,\delta_{j,j_1} \Big]
  \,\frac{\spb1.3\,\spaa2.\ell.(1+2+3).2}{\spa2.3\,s_{12}\,s_{123}
}\nonumber \\
 &-&\Big[(t^{c_3})_{j j_1}\,\delta_{j_2,\jb}-\frac{1}{N}\,(t^{c_3})_{j_2 j_1}\,\delta_{j,\jb} \Big]
  \,\frac{\spaa2.\ell.(1+2+3).2}{\spa1.3\, \spa2.3\,s_{123} }\nonumber \\
 &+&\Big[(t^{c_3})_{j j_1}\,\delta_{j_2,\jb}-\frac{1}{N}\,(t^{c_3})_{j \jb}\,\delta_{j_2,j_1} \Big]
  \,\frac{\spab2.{\ell_{123}}.1\,\spab2.{\ell}.3}{\spa2.3\,s_{12}\,D(\ell_{3})}\nonumber \\
 &+&\Big[(t^{c_3})_{j_2 \jb}\,\delta_{j,j_1}-\frac{1}{N}\,(t^{c_3})_{j \jb}\,\delta_{j_2,j_1} \Big]
  \,\frac{\spab2.\ell.1\,\spab2.{\ell_{123}}.3}{\spa2.3\,s_{12}\,D(\ell_{12})} \Bigg\}
\end{eqnarray}
\begin{eqnarray}
{\cA}^{{\qb}qg}_3(\ell;1_{\qb}^+,2_{q}^-, 3^-_g;\ellb) & = &
 i \sqrt{2} g_s^3\, \Bigg\{
      -\Big[(t^{c_3})_{j j_1}\,\delta_{j_2,\jb}-(t^{c_3})_{j_2 \jb}\,\delta_{j,j_1} \Big]
       \,\frac{\spa2.3\,\spbb1.\ell.(1+2+3).1}{\spb1.3\,s_{12}\,s_{123}}\nonumber \\
     &-&\Big[(t^{c_3})_{j_2 \jb}\,\delta_{j,j_1}-\frac{1}{N}\,(t^{c_3})_{j_2 j_1}\,\delta_{j,\jb} \Big]
       \,\frac{\spbb1.\ell.(1+2+3).1}{\spb1.3\,\spb2.3\,s_{123}}\nonumber \\
     &-&\Big[(t^{c_3})_{j j_1}\,\delta_{j_2,\jb}-\frac{1}{N}\,(t^{c_3})_{j \jb}\,\delta_{j_2,j_1} \Big]
        \,\frac{\spab2.{\ell_{123}}.1\spba1.{\ell}.3}{\spb1.3\,s_{12}\,D(\ell_{3})}\nonumber \\
     &-&\Big[(t^{c_3})_{j_2 \jb}\,\delta_{j,j_1}-\frac{1}{N}\,(t^{c_3})_{j \jb}\,\delta_{j_2,j_1} \Big]
      \,\frac{\spab2.\ell.1\,\spba1.{\ell_{123}}.3}{\spb1.3\,s_{12}\,D(\ell_{12})}  \Bigg\}\nonumber \\
\end{eqnarray}
Useful results for tree graph amplitudes with two massive quarks and $n$-gluons for amplitudes (a) with
all gluon helicities the same, or (b) with all but one helicities the same, have been given in ref.~\cite{Ochirov:2018uyq}.

\section{Numerical value of coefficients at a given phase-space point}
In order to assist in the reconstruction of the coefficients 
in a numerical program we give the value of all the needed 
base set of integral coefficients at a given data point.
The point we choose is given by (with $p = (E,p_x,p_y,p_z)$),
\begin{eqnarray} \label{kinpoint_paper}
p_1      &=&(-15\kappa,-10\kappa, +11\kappa,  +2\kappa)\nonumber \\
p_2      &=&(-9\kappa,  +8\kappa,  +1\kappa,  -4\kappa)\nonumber \\
p_3      &=&(-21\kappa, +4\kappa, -13\kappa, +16\kappa)\nonumber \\
p_4      &=&(-7\kappa,  +2\kappa,  -6\kappa,  +3\kappa)\nonumber \\
p_\Higgs &=&(+52\kappa, -4\kappa,  +7\kappa, -17\kappa)
\end{eqnarray}
with $\kappa=1/\sqrt{94}$~GeV and $p_\Higgs=-p_1-p_2-p_3-p_4$.
This fixes $s_{1234}=25$~GeV$^2$, $M_\Higgs=5$~GeV and we further choose $m=1.5$~GeV.
The numerical values of the coefficients and the rational terms are given in 
Tables~\ref{CoeffValuesAllPlusOneMinus},~\ref{CoeffValuesTwoMinus} and \ref{aqggtable}.

\begin{table}
\begin{center}
\begin{tabular}{|l|l|l|l|l|}
\hline
Helicities & Coefficient        & Real Part     & Imaginary Part & Absolute Value \\
\hline
+\,+\,+\,+ &  $d_{1\x2\x34}$  &     1.7494424584 &     0.9145464014 &     1.9740678903 \\
           &  $d_{1\x23\x4}$  &     5.9642590946 &     8.6101701300 &    10.4741308095 \\
           &  $d_{1\x2\x3}$   &    11.9903841330 &    27.4958119676 &    29.9964829174 \\
           &  $c_{1\x234}$    &    18.9100021625 &    56.3253050259 &    59.4148817052 \\
           &  $r$             &    -6.4366316747 &   -19.1721417745 &    20.2237792595 \\
\hline
+\,+\,+\,-- &  $d_{1\x2\x34}$  &   -24.3908884307 &   -34.1026538098 &    41.9273948071 \\
           &  $d_{1\x4\x32}$  &   -22.2037730441 &   -29.7427434881 &    37.1165505885 \\
           &  $d_{2\x1\x43}$  &    16.2217246906 &   -62.9923572563 &    65.0475320412 \\
           &  $d_{2\x34\x1}$  &   -66.4392574700 &    12.9335349956 &    67.6864185834 \\
           &  $d_{4\x3\x21}$  &     8.2313626631 &    -0.7960661671 &     8.2697673869 \\
           &  $d_{1\x23\x4}$  &     2.0815256682 &    -1.7746340633 &     2.7353382179 \\
           &  $d_{2\x3\x4}$   &    -0.9920798783 &    -1.5084323993 &     1.8054336843 \\
           &  $d_{1\x2\x3}$   &    22.2417370205 &     1.7066361068 &    22.3071170816 \\
           &  $d_{3\x4\x1}$   &     0.8741489856 &    -5.3830902459 &     5.4536040418 \\
           &  $c_{3\x4}$      &    -0.0041638038 &     0.0115576710 &     0.0122848289 \\
           &  $c_{2\x34}$     &     3.1035163815 &    -0.1080335333 &     3.1053961381 \\
           &  $c_{1\x43}$     &     6.9656648763 &    -0.8139894264 &     7.0130639492 \\
           &  $c_{4\x123}$    &    12.7875856866 &     2.0711796271 &    12.9542322327 \\
           &  $c_{1\x234}$    &   -41.8343835373 &   -39.3169799861 &    57.4102827129 \\
           &  $c_{2\x341}$    &    -0.0578594858 &   -18.9964204402 &    18.9965085545 \\
           &  $c_{12\x34}$    &    12.4596639704 &   -35.5399553316 &    37.6607441672 \\
           &  $b_{34}$        &    -0.0409808246 &     0.1015477837 &     0.1095051613 \\
           &  $b_{234}$       &     0.2936947594 &    -0.0490382211 &     0.2977605730 \\
           &  $b_{1234}$      &    -0.9341272666 &    -0.1920882562 &     0.9536727156 \\
           &  $r$             &    -3.8872487587 &    10.3025699409 &    11.0115235230 \\
\hline
\end{tabular}
\caption{Numerical values of coefficients and rational terms for +\,+\,+\,+ and +\,+\,+\,-- helicities
of of $gggg\Higgs$ at kinematic point, \ref{kinpoint_paper}.}
\label{CoeffValuesAllPlusOneMinus}
\end{center}
\end{table}

\begin{table}
\begin{center}
\begin{tabular}{|l|l|l|l|l|}
\hline
Helicities & Coefficient        & Real Part     & Imaginary Part & Absolute Value \\
\hline
+\,--\,+\,-- &  $d_{4\x3\x21}$  &    -7.6953556408 &    -6.4085129013 &    10.0143664825 \\
           &  $d_{1\x23\x4}$  &    -2.3752436126 &     1.8890031582 &     3.0348171528 \\
           &  $d_{1\x2\x3}$   &   -14.9620839628 &   -39.7624750054 &    42.4843309359 \\
           &  $c_{3\x4}$      &    -0.0785670511 &     0.1424216217 &     0.1626551563 \\
           &  $c_{2\x34}$     &     5.9965313107 &    -4.5593453199 &     7.5329952546 \\
           &  $c_{12\x34}$    &   -22.6495761599 &    21.1031361652 &    30.9571584004 \\
           &  $c_{1\x234}$    &    -9.2908649214 &    -2.0570320613 &     9.5158579166 \\
           &  $b_{34}$        &     0.0682767006 &     0.0433975227 &     0.0809015007 \\
           &  $b_{234}$       &     2.3825660060 &    -0.8884219110 &     2.5428162074 \\
           &  $b_{1234}$      &    -4.2679248420 &     3.0120566624 &     5.2237599288 \\
           &  $r$             &     2.3606586680 &     0.8025702116 &     2.4933568319 \\
\hline
+\,+\,--\,-- &  $d_{1\x2\x34}$  &    -0.0125227093 &    -0.5196169994 &     0.5197678754 \\
           &  $d_{1\x4\x32}$  &     8.4132295790 &  -459.7920528912 &   459.8690186715 \\
           &  $d_{2\x34\x1}$  &    62.3890431832 &   -51.7566409441 &    81.0625844095 \\
           &  $d_{1\x23\x4}$  &    -3.4240833144 &     4.6410747884 &     5.7674883386 \\
           &  $d_{1\x2\x3}$   &    -5.4385640586 &    -6.5811803202 &     8.5375589853 \\
           &  $c_{2\x3}$      &   -41.9249189131 &   -23.3819669075 &    48.0043248295 \\
           &  $c_{1\x23}$     & -1036.7850502032 &  -480.4884415677 &  1142.7127297817 \\
           &  $c_{23\x41}$    &  1080.7316959848 &   740.3414428401 &  1309.9948284984 \\
           &  $c_{1\x234}$    &    22.0281339875 &  -305.0638529285 &   305.8581256899 \\
           &  $b_{23}$        &    -5.3092820284 &    -9.1916846550 &    10.6148736429 \\
           &  $b_{234}$       &    -0.8234906782 &     5.4869406052 &     5.5483920285 \\
           &  $b_{1234}$      &    26.7533037866 &    15.0643874405 &    30.7030134100 \\
           &  $r$             &     1.3340510134 &     0.8053633104 &     1.5583010518 \\
\hline
\end{tabular}
\caption{Numerical values of coefficients and rational terms for +\,--\,+\,-- and +\,+\,--\,-- helicities
of $gggg\Higgs$ at kinematic point, \ref{kinpoint_paper}.}
\label{CoeffValuesTwoMinus}
\end{center}
\end{table}

 \begin{table}
 \begin{center}
 \begin{tabular}{|l|l|l|l|l|}
 \hline
 Helicities & Coefficient & Real Part & Imaginary Part & Absolute Value \\
 \hline
+\,--\,+\,+ &$d_{4\x3\x21}$&    4.0685161820&   -4.0500901147&    5.7407363517\\
            &$d_{4\x21\x3}$&  425.5033072909& 1294.6650310348& 1362.7951449502\\
            &$c_{3\x21}$   &  -73.7590711176& -242.0029027576&  252.9936867102\\
            &$c_{12\x34}$&     14.7023801790&   -6.9563781545&   16.2650293561\\
            &$c_{4\x123}$&     32.6756691373&   92.0151860555&   97.6447326711\\
            &$c_{3\x412}$&     87.6696567417&  249.0681532719&  264.0471807982\\
            &$b_{12}$&          1.4197098901&    0.3520351648&    1.4627046624\\
             &$b_{124}$&       -0.5233075583&   -1.0058683561&    1.1338527023\\
            &$b_{123}$&        -0.7954324907&   -2.6428744805&    2.7599815882\\
            &$b_{1234}$&       -0.1009698411&    3.2967076718&    3.2982535351\\
            &$r$&              -5.4119652752&    0.7153882121&    5.4590428130\\
 \hline
+\,--\,--\,+&$d_{4\x3\x21}$&    1.0782715488&   -4.7280903169&    4.8494852900\\
            &$d_{4\x21\x3}$&   13.3402061977&   -3.4340877490&   13.7751246842\\
            &$c_{3\x21}$   &   -0.8475265952&    2.4513967233&    2.5937708504\\
            &$c_{3\x4}$    &    0.0094395944&    0.0336984573&    0.0349955993\\
            &$c_{12\x34}$&     -3.7289304305&  -10.8894201371&   11.5101864919\\
            &$c_{4\x123}$&      1.7886984296&   -1.6881229718&    2.4595123988\\
            &$c_{3\x412}$&      1.8223566626&    0.0772529014&    1.8239933708\\
            &$b_{12}$&         -0.1514791222&   -0.2049070241&    0.2548191770\\
            &$b_{34}$&          0.0053272676&    0.1024178437&    0.1025562991\\
             &$b_{124}$&        0.2906270969&   -0.1175213822&    0.3134890504\\
            &$b_{123}$&         0.6201039140&    0.1805231188&    0.6458463135\\
            &$b_{1234}$&       -0.7645791563&    0.0394874437&    0.7655981613\\
            &$r$&              -0.2305425036&    1.4755649890&    1.4934663983\\
 \hline
+\,--\,+\,--&$c_{4\x123}$&      0.0795879764&   -1.9432491013&    1.9448782264\\
            &$b_{123}$&        -1.0956687877&    0.8092787161&    1.3621388082\\
 \hline
 \end{tabular}
 \caption{Numerical values of the needed coefficients
  and rational terms for +\,--\,+\,+, +\,--\,--\,+
  and +\,--\,+\,-- helicities
 of the $\qb q gg\Higgs$ process at kinematic point, \ref{kinpoint_paper}.}
 \label{aqggtable}
 \end{center}
 \end{table}

In addition the values of the colour-ordered amplitudes after substitution of
all the scalar integrals and including the rational terms are,
\begin{align}
H^{1234}(1^+,2^+,3^+,4^+;\Higgs)&=+29.24088185 - 46.63892079\, i,
 \; & |H^{1234}(1^+,2^+,3^+,4^+;\Higgs)| &= 55.04741687 \nonumber \\
H^{1234}(1^+,2^+,3^+,4^-;\Higgs)&=-28.10008864 + 9.836858255\, i,
 \; & |H^{1234}(1^+,2^+,3^+,4^-;\Higgs)| &= 29.77211383 \nonumber \\
H^{1234}(1^+,2^-,3^+,4^-;\Higgs)&=+4.580787288 + 7.498254006\, i,
 \; & |H^{1234}(1^+,2^-,3^+,4^-;\Higgs)| &= 8.786775593 \nonumber \\
H^{1234}(1^+,2^+,3^-,4^-;\Higgs)&=+0.369177073-1.815728344\, i,
 \; & |H^{1234}(1^+,2^+,3^-,4^-;\Higgs)| &= 1.852879146 \nonumber \\
\end{align}
\begin{align}
H^{34}(1^+,2^-,3^+,4^+;\Higgs)&= -8.998 796 972-13.029 709 81 \,i, \; 
&|H^{34}(1^+,2^-,3^+,4^+;\Higgs)|=15.83514081 \nonumber \\
H^{34}(1^+,2^-,3^-,4^+;\Higgs)&= -3.850 947 633 +1.791 151 530\,i, \;
&|H^{34}(1^+,2^-,3^-,4^+;\Higgs)|=4.247 119 197 \nonumber \\
H^{34}(1^+,2^-,3^+,4^-;\Higgs)&= -0.412 185 752+ 7.682 564 596 \,i, \; 
&|H^{34}(1^+,2^-,3^+,4^-;\Higgs)|=7.693 613 966  \nonumber \\
\end{align}
\begin{align}
H^{4q}(1^{+},2^{-},3^{+},4^{-};\Higgs) &= 0.620 045 806+4.703 0845 622 \,i, \; 
&|H^{4q}(1^{+},2^{-},3^{+},4^{-};\Higgs)|=4.743 781 319  \nonumber \\
\end{align}

\bibliography{notes}
\bibliographystyle{JHEP}
\end{document}